\documentclass[aps,showpacs,twocolumn,twoside,amsmath,amssymb,groupedaddress,superscriptaddress,prc]{revtex4-2}
\usepackage{multirow}
\usepackage{tikz}
\usepackage{epstopdf}
\usepackage[percent]{overpic}  
\usepackage{scrextend}  
\usepackage{xcolor,xspace}
\usepackage[ulem=normalem]{changes}
\usepackage{hyperref}
\hypersetup{colorlinks=True,urlcolor=blue,linkcolor=blue,citecolor=blue,filecolor=black}

\colorlet{Changes@Color}{red}  
\usepackage{amsmath,bm}
\usepackage{dcolumn,color,footnote,braket}
\usepackage{url,longtable,tabularx}
\usepackage{fancybox}
\usetikzlibrary{matrix}

\begin{document}

\title{
Hexadecapole axial collectivity in the rare earth region, a beyond mean field study}

\author{C.V. Nithish Kumar}
\email{nithishkumarcv@gmail.com}
\affiliation{Departamento de F\'\i sica Te\'orica and CIAFF, Universidad
Aut\'onoma de Madrid, E-28049 Madrid, Spain}

\author{L.~M.~Robledo}
\email{luis.robledo@uam.es}
\affiliation{Departamento de F\'\i sica Te\'orica and CIAFF, Universidad
Aut\'onoma de Madrid, E-28049 Madrid, Spain}
\affiliation{Center for Computational Simulation,
Universidad Polit\'ecnica de Madrid,
Campus de Montegancedo, Bohadilla del Monte, E-28660-Madrid, Spain
}

\date{\today}

\begin{abstract}
Hexadecapole collectivity and its interplay with quadrupole degrees 
of freedom is studied in an axial symmetry preserving framework based
on the Hartree Fock Bogoliubov (HFB) plus generator coordinate method (GCM).
Results are obtained for several even-even isotopes of Sm and Gd with 
various parametrizations of the Gogny force. The analysis of the results
indicates the strong coupling between the quadrupole and hexadecapole 
degrees of freedom. The first two excited states are vibrational in 
character in most of the cases. The impact of prolate-oblate shape mixing in the
properties of hexadecapole states is analyzed. 
\end{abstract}

\maketitle

\section{Introduction}

Understanding the impact of the intrinsic shape of nuclei in the 
dynamics of their lowest lying collective states  is one of the most 
important challenges in nuclear structure nowadays. To quantify the 
intrinsic shape of the nucleus, multipole moments of the matter 
distribution are introduced; of which  the quadrupole moment is the 
most important one. Moreover, multipole moments are also used as 
collective variables in order to characterize collective dynamics. The 
presence of non-zero  multipole moments, signaling whether a nucleus is 
deformed or not, influence  properties of the collective  spectrum such 
as rotational bands, parity doublets, etc. On the other hand, dynamical 
deformation, associated with vibrations around the equilibrium position 
determine the properties of the so-called $\beta$ and $\gamma$ bands in 
the quadrupole case. These ideas can be extended further to higher 
order multipole excitations like the celebrated $3^{-}$ octupole 
vibrational state in $^{208}$Pb. Fluctuations on the collective shape 
degrees of freedom around the ground state equilibrium point can be 
analyzed in terms of collective wave functions. These obtained through 
well-defined theoretical procedures like the generator coordinate 
method (GCM) based on Hartree Fock Bogoliubov (HFB) mean field wave 
functions. 

By looking at the energy as a function of the relevant quadrupole 
deformation parameters obtained in self-consistent mean field 
calculations one can introduce important concepts characterizing the 
nucleus, like prolate/oblate ground states, triaxiallity, shape 
coexistence, etc. Also important is the negative parity set of octupole 
moments. They carry three units of angular momentum and  negative 
parity. Therefore they are disconnected from the quadrupole degrees of 
freedom except in nuclei breaking reflection symmetry in their ground 
state. This is a direct consequence of the different parity quantum 
number associated with the two sets. Therefore, it is to be expected 
that the next shape multipole moment to strongly couple to the 
quadrupole one is the positive parity hexadecapole moment carrying four 
units of angular momentum. Many different kinds of calculations predict 
permanent hexadecapole deformation in several regions of the nuclear 
chart \cite{Hilaire2005,Moeller2016,Scamps2021,Lalazissis1999}. Ground 
state deformation has mostly $K=0^{+}$ character and the sign of the 
associated deformation parameter $\beta_4$ determines whether the 
nucleus has an equilibrium ``square-like" shape ($\beta_{4} <0$) or a 
``diamond-like" ($\beta_{4} >0$). On the other hand, hexadecapole 
$K=4^{+}$ vibrational bands, analogous to the $\gamma$ bands of the 
quadrupole dynamics, have been identified experimentally -- see 
\cite{Garrett2005,Phillips2010,Hartley2020} for recent examples. 
Considering $K=4^{+}$ hexadecapole bands implies also considering the 
coupling with $K=2^{+}$ and $K=0^{+}$ bands \cite{Magierski1995} which 
implies a GCM calculation with five degrees of freedom (three 
hexadecapole and two quadrupole) which is out of reach with present day 
available computational capabilities. This is one of the reasons why we 
focus as a first step on the axially symmetric hexadecapole degree of 
freedom associated with $Q_{40}$. We will analyze its impact on the 
binding energy gain as well as the energy of the hexadecapole 
$\beta_{4}$-vibration-like excitation. 

For $K=0^{+}$ states the $\beta$ quadrupole deformation parameter is 
expected to be dominant degree of freedom. In this case,  the energy as 
a function of the $Q_{40}$ hexadecapole  deformation parameter should 
be parabolic and the $\beta_{4}$ zero point energy of collective motion 
cancels out the zero point energy correction leaving the energy 
unaffected. Contrary to this expectation, the results of our 
calculations show that the consideration of $\beta_{4}$ in the ground 
state dynamic increases in some cases the binding energy by around 
500-600 keV, a quantity that is similar to the one gained by including 
the quadrupole degree of freedom as discussed below.

Recently, it has been argued that hexadecapole deformation can leave 
its imprint in the elliptic flow of particles in relativistic 
collisions of $^{238}$U nuclei at the BNL Relativistic Heavy Ion 
Collider (RHIC) \cite{ryssens2023}. Therefore, it is of considerable 
interest to analyze the impact of dynamical fluctuations in the 
hexadecapole properties of the target nuclei. 

There are several examples in the rare-earth region of nuclei with a 
large number of excited $0^{+}$ states at low energies (typically below 
3 MeV) that are not easy to interpret \cite{Meyer2005}. It has been 
argued that $\beta$ vibration could be one of these states. Other 
candidates could be a double phonon excitation. One can also argue that 
a $\beta_{4}$ vibrational state could be found among that large number 
of $0^{+}$ states. As discussed below, this possibility is largely 
suppressed due to the high excitation energy predicted for this states. 

Last but not least, hexadecapole deformation can play a role in the 
value of the neutrino-less double beta decay nuclear matrix element in 
nuclei in the rare-earth region around $^{150}$Nd \cite{Engel2017}.

In this paper the combined dynamic of the quadrupole and hexadecapole 
$K=0$ collective degrees of freedom is analyzed with a theoretical 
framework based on the GCM built on top of a set of HFB mean field wave 
functions. As the HFB systematic with the Gogny D1S shows (see below) 
the region with the largest ground state $\beta_{4}$ values is located 
in the nuclear chart at around $Z=64$ and $N=90$. For this reason, the 
nuclei chosen for the present study are several isotopes of Sm ($Z=62$) 
and Gd ($Z=64$). 

%

\section{Theoretical method \label{sec:method}}

As a first step, we carry out self-consistent mean field calculations 
with the finite range Gogny force in order to obtain a set of HFB wave 
functions $|\varphi (\beta_{2}, \beta_{4})\rangle$ satisfying 
constraints on the quadrupole $Q_{20}$ and hexadecapole $Q_{40}$ 
moments. In order to have a description independent of mass number, we 
will parameterize the moments in terms of the $\beta_{l}$ deformation 
parameters \cite{Egido1992}
\begin{equation} \label{betal}
\beta_{l} = \frac{\sqrt{4 \pi (2l+1)}}{3 R_{0}^{l} A} Q_{l0}
\end{equation}
where, $R_{0} = 1.2 A^{1/3}$~fm and $A$ is the mass number. As it is 
customary in calculations with the Gogny force we have expanded the 
Bogoliubov quasiparticle operators in a harmonic oscillator (HO) basis. 
The optimal number of HO shells to be used for a given nucleus depends 
on its mass number as well as the variety of shapes to be considered. 
We have taken 17 major shells in the present study involving rare earth 
nuclei and checked that the results do not change in a significant way 
(exception made of a slight increase in binding energy) when the 
calculation is repeated with 19 major shells. More important is the 
fact that all the wave functions to be used in the subsequent generator 
coordinate method (GCM) calculation, must have the same oscillator 
lengths to avoid problems with the traditional formulas in the 
evaluation of the operator overlaps required by the GCM 
\cite{Rob94,Robledo2022,Robledo2022a}. The specific value of the 
oscillator lengths is rather irrelevant given the huge basis size used. 
We have chosen equal oscillator lengths $b_{\perp}=b_{z}$ and for its 
value the $b=1.01 A^{1/6}$ estimation. The set of HFB wave functions  
enter linear combinations with weights $f_{\sigma}(\beta_{2}, 
\beta_{4})$
\begin{equation}
|\Psi_{\sigma}\rangle = \int d\beta_{2} d\beta_{4} f_{\sigma}(\beta_{2}, \beta_{4}) |\varphi (\beta_{2}, \beta_{4})\rangle
\end{equation}
defining the set of physic states $|\Psi_{\sigma}\rangle$ labeled by the
$\sigma$ quantum number. The $f_{\sigma}$ amplitudes are determined by
the Ritz variational principle on the energy and are the solution of the
Griffin-Hill-Wheeler (GHW) equation 
\begin{equation} \label{HW-equation}
\int d \bm{\beta}'
\left(
{\cal{H}}(\bm{\beta}, \bm{\beta}') - E_{\sigma}^{\pi}
{\cal{N}}(\bm{\beta}, \bm{\beta}')
\right)
f_{\sigma}^{\pi} (\bm{\beta}') = 0
\end{equation} 
where the shorthand notation $\bm{\beta}=(\beta_{2}, \beta_{4})$ has been
introduced. The Hamiltonian and norm kernels are given by
\begin{align} \label{GCM-PROJEDF-hnk}
{\cal{H}}(\bm{\beta}, \bm{\beta}') &=
\langle {\varphi} (\bm{\beta}) | 
\hat{H} [\rho^{GCM}(\vec{r}) ] | {\varphi} (\bm{\beta}') \rangle
\nonumber\\
{\cal{N}}(\bm{\beta}, \bm{\beta}') 
&= \langle {\varphi} (\bm{\beta}) |  {\varphi} (\bm{\beta}') \rangle
\end{align}
where we have used the ``mixed" density prescription 
$\rho^{GCM}(\vec{r})$ for the density dependent term of the Hamiltonian 
(see, Refs.~\cite{Robledo2010,Sheikh2021} for a discussion of the 
associated problematic). We also include a perturbative correction  in 
the Hamiltonian  kernel ${\cal{H}}(\bm{\beta}, \bm{\beta}')$ to correct 
for deviations in both the proton and neutron numbers \cite{RodriguezGuzman2012}. 

\begin{figure}
\includegraphics[angle=0,width=0.95\columnwidth]{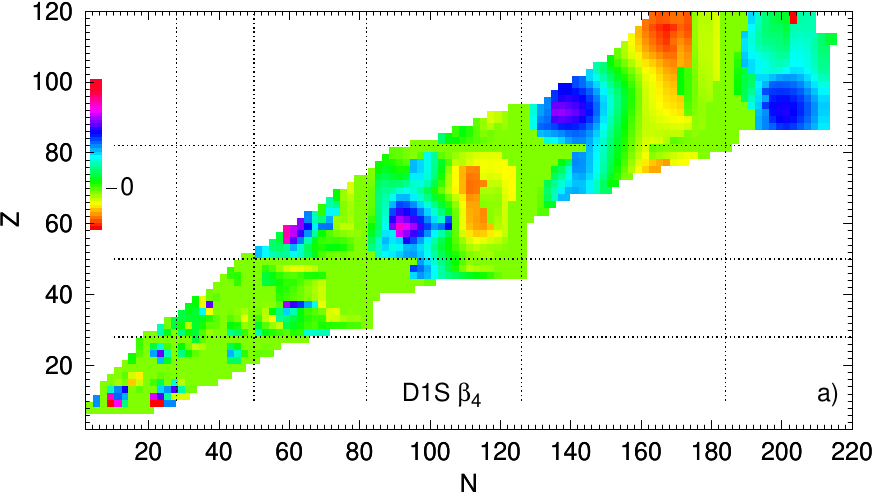}  
\caption{(Color online) Color map of the ground state hexadecapole
deformation obtained with the HFB method and the Gogny D1S force. The
color scale ranges from $\beta_{4}=-0.2$ up to $\beta_{4}=0.5$ with
$\beta_{4}=0$ corresponding to green.   
}
\label{Beta4_MAP} 
\end{figure}

Since the wave functions  $| {\varphi} (\bm{\beta}) \rangle$ do not form an orthonormal set, 
the $f_{\sigma}^{\pi} (\bm{\beta})$  
are not probability amplitudes. One can define genuine probabilities by folding them
with a square root of the norm ${\cal{N}}^{\frac{1}{2}}$ kernel
\begin{equation} \label{cll-wfs-HW} 
G_{\sigma}^{\pi} (\bm{\beta}) =   \int d \bm{\beta} {\cal
{N}}^{\frac{1}{2}} (\bm{\beta}, \bm{\beta}')  f_{\sigma}^{\pi}(\bm{\beta}')
\end{equation}  
See, Refs. \cite{Robledo2019} for details on how to solve the GHW equation
and how to interpret its solution. 

%

As it is customary in this type of calculations the integrals over the 
continuous $\beta_{2}$ and $\beta_{4}$ variables are discretised in a 
mesh with step sizes $\Delta\beta_{2}=0.02$ and $\Delta\beta_{4}=0.02$. 
The intervals considered are $[-0.4,0.8]$ for $\beta_{2}$ and 
$[-0.4,0.6]$ for $\beta_{4}$. We have checked that reducing the number 
of point in each direction to half the nominal value has a negligible 
impact on the results. 

For the calculations presented in this study we have used two sets of 
parametrizations of the Gogny force. One is the traditional D1S 
parametrization \cite{berger1984} which has been used for more than 
forty years to describe many nuclear properties all over the Segr\`e 
chart. The other one is the recently proposed D1M* 
\cite{GonzalezBoquera2018} parametrization which is a variation of D1M 
\cite{Goriely2009} retaining most of its properties but improving on 
the description of neutron stars by imposing a different value of the 
slope of the symmetry energy \cite{Vinas2019}. 

%

\section{Results and discussions\label{sec:results}}

Results obtained for two isotopic chains in the rare earth region are 
discussed in this section. The nuclei considered are those of Sm 
($Z=62$) and Gd ($Z=64$) and the choice is guided by the results of 
systematic HFB calculations with the Gogny D1S force. In Fig.
\ref{Beta4_MAP} the ground state $\beta_{4}$ deformation parameter 
obtained  for a large set 
of even-even nuclei is depicted as a color map.  The ground state 
$\beta_{4}$ values in the figure range from $\beta_{4}=-0.1$ up to 
$\beta_{4}=0.3$. One observes that a large fraction of nuclei show zero 
hexadecapole deformation in their ground state. The largest positive 
values are obtained in the lower $Z$ sector of rare earth (actinide) 
with proton numbers 60 (90) and neutron numbers 90 (136), a few units 
above magic numbers. On the other hand, the largest negative values are 
also located in the same regions but this time in the upper $Z$ sector 
with values around 72  (118) and neutron numbers around 110 (170) which 
are a few units below magic numbers. There are two additional regions 
with large positive $\beta_{4}$ values at the proton drip line with 
$Z=60$ and close to the neutron drip line at $Z=90$ and $N=200$. The 
regions of positive and negative $\beta_{4}$ values are consistent with 
the polar-gap model of Ref.~\cite{Bertsch1968} developed to understand 
$\beta_{4}$ deformation parameters in the rare earth nuclei 
\cite{Hendrie1968}. In the model, positive (negative) $\beta_{4}$ 
values appear at the beginning (end) of the shell. 

The figure points to very large positive $\beta_{4}$ deformation 
parameters in the region under study with N around 90 and Z around 60. 
For the nucleus $^{154}$Sm considered below a ground state hexadecapole 
deformation $\beta_{4}=0.21$ is obtained. The $\beta_{4}$ deformation 
parameter is obtained with Eq. (\ref{betal}) and may differ from other 
deformation parameters defined, for instance, in terms of $\langle 
r^{4}\rangle$ instead of $R_{0}^{4}$. Those tend to be smaller, 
and in the case of $^{154}$Sm one gets $\beta_{4}=0.17$ instead of 
$\beta_{4}=0.21$ obtained with  Eq. (\ref{betal}).

Finally, let us mention that our results for $\beta_{4}$  are 
consistent in absolute value with those of a recent Skyrme interaction 
BSkG1 \cite{Scamps2021}. However, both our results and the ones in 
\cite{Scamps2021} tend to be larger than the ones obtained with mic-mac 
models \cite{Moeller2016}. For instance, for $^{154}$Sm Moller et al 
obtain $\beta_{4}=0.11$. The source for the discrepancy could be 
associated to the different definition of the $\beta_{l}$ parameters as
discussed above.
 
In a recent publication  \cite{Spieker2023} very large values of both 
$\beta_{2}$ and $\beta_{4}$ have been obtained in inelastic proton 
scattering experiments in inverse kinematics on the rare isotopes 
$^{74}$Kr and $^{76}$Kr. For $^{76}$Kr one obtains $\beta_{2}=0.40$ and 
$\beta_{4}=0.201$ whereas for $^{74}$Kr one obtains $\beta_{2}=0.35$ 
and $\beta_{4}=0.23$. Those findings do not agree with the results 
obtained with Gogny D1S (see Fig. \ref{Beta4_MAP} above and Ref. 
\cite{Hilaire2005}) that seem to favor spherical or nearly spherical 
ground states for those two isotopes. The discrepancy could possibly be resolved
by taking into account that the Kr isotopes represent one of the 
most prominent examples of shape coexistence with very flat potential 
energy surfaces and large fluctuations along the quadrupole degree of 
freedom. This is the realm where the theoretical techniques used in 
this paper are most relevant and therefore the present calculations are 
being extended to the Kr region and will be reported in the future. 

\begin{figure}
\includegraphics[angle=0,width=0.95\columnwidth]{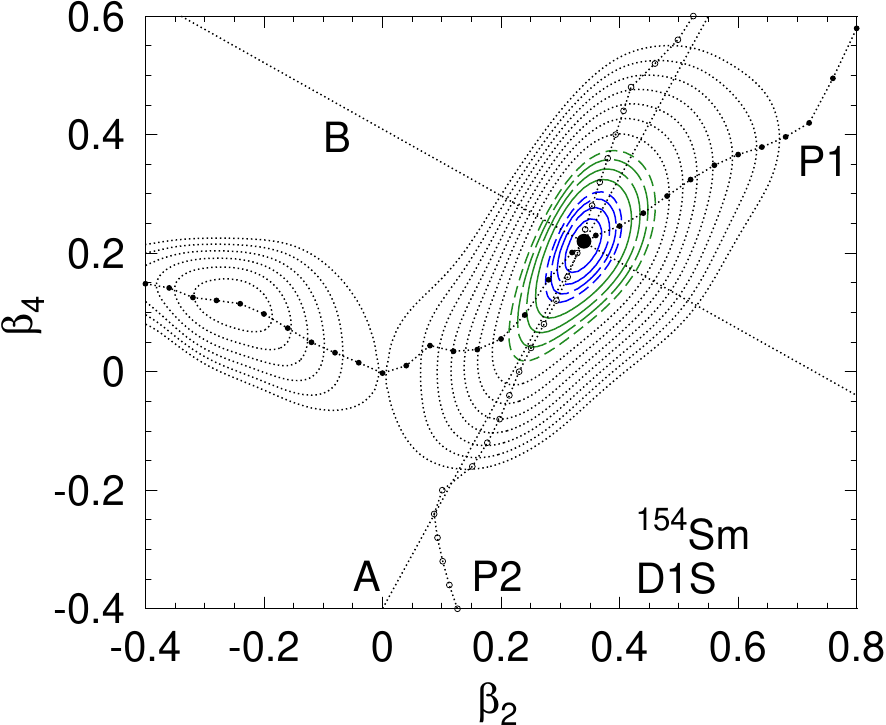}  
\caption{(Color online) Contour plot of the HFB energy as a function of 
$\beta_{2}$ (horizontal axis) and $\beta_{4}$ (vertical axis). The 
position of the absolute minimum is indicated by a large dot. The lowest four 
contours (blue) correspond to energies $E_{min}+\delta E n$ with 
$\delta E=0.25$ MeV. The next four (green) correspond to energies 
$E_{min}+ 1 \textrm{MeV} + \delta E n$ with $\delta E=0.5$ MeV. The 
remaining contours, starting at $E_{min} + 4 \textrm{MeV} $ are 
separated by 1 MeV. The filled (empty) dots connected by a curve 
correspond to the self-consistent values of $\beta_{4}$ ($\beta_{2}$) 
obtained in the 1D calculation as a function of $\beta_{2}$ 
($\beta_{4}$). The paths are labeled P1 and P2, respectively. 
The two perpendicular dotted lines crossing at the 
minimum are drawn along the two principal axes (A and B) of the 
parabola that approximates the HFB energy around the minimum.
}
\label{154SmEB2B4} 
\end{figure}

\begin{figure}
\includegraphics[angle=0,width=0.95\columnwidth]{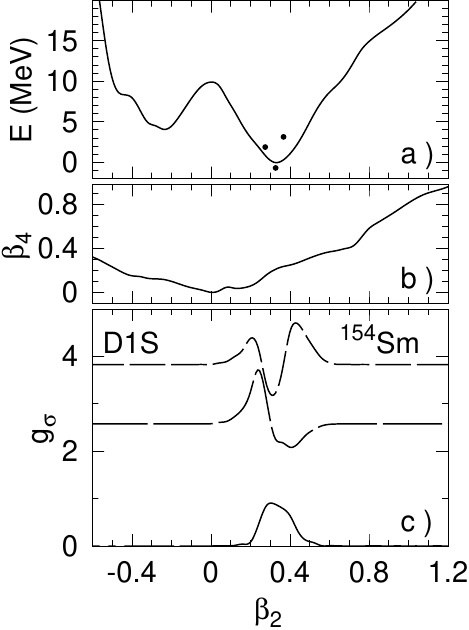}  
\caption{(Color online) In panel a) the potential energy surface as a function
of $\beta_{2}$ is drawn for the nucleus $^{154}$Sm along the path P1 of Fig.~\ref{154SmEB2B4}. 
The three dots correspond
to the three lowest solutions of the 1D GCM and are plotted at the corresponding
energies and average $\beta_{2}$ values. In panel b) the self-consistent $\beta_{4}$
deformation is plotted as a function of $\beta_{2}$. Finally, in panel c) the
collective amplitudes of the GCM $g_{\sigma} (\beta_{2})$ are plotted for 
the lowest three solutions of the GCM. The curves are shifted by the
excitation energy of the corresponding states (y-axis).  
}
\label{154Sm1D_P1} 
\end{figure}

\subsection{The nucleus $^{154}$Sm\label{sec:pes}}

\begin{figure}
\includegraphics[angle=0,width=0.95\columnwidth]{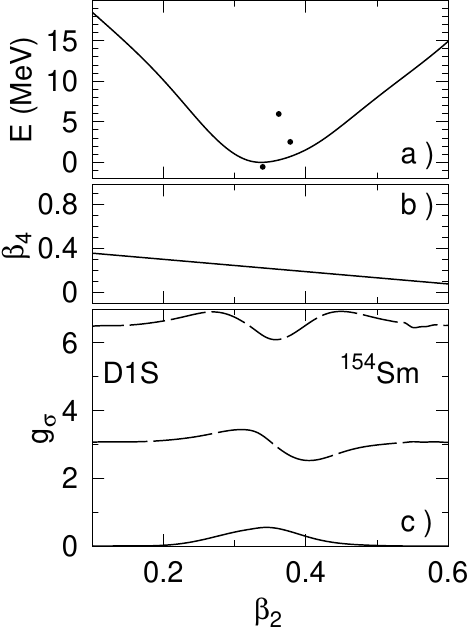}  
\caption{(Color online) Same as Fig.~\ref{154Sm1D_P1} but for the path B shown
in Fig.~\ref{154SmEB2B4}.  
}
\label{154Sm1D_B} 
\end{figure}

In this section we discuss at length all the details and peculiarities 
of our methodology in the paradigmatic case of $^{154}$Sm.  The HFB 
energy corresponding to the nucleus $^{154}$Sm obtained with the D1S 
parametrization of the Gogny force is shown in Fig. \ref{154SmEB2B4}. 
The energy shows a parabolic behavior around the minimum (marked by a 
large dot) with principal axes going in directions not parallel to the 
horizontal and vertical axes. This fact implies that in this case 
$\beta_{4}$ changes substantially when one moves along the bottom of 
the energy valley as a function of $\beta_{2}$. In a more quantitative 
way we can say that the hexadecapole deformation corresponding to the 
bottom of the energy valley shows an approximate linear relation 
$\beta_{4}= 1.8 \beta_{2}-0.4$ as a function of $\beta_{2}$. This 
direction in the $\beta_{2}-\beta_{4}$ plane is denoted as A. The 
perpendicular direction, denoted by B, will be discussed below. The 
linear behavior implies that a GCM calculation using as generating 
parameter the $\beta_{4}$ deformation alone (path P2 in the figure) 
will explore roughly the same configuration around the energy minimum 
as a GCM calculation with the $\beta_{2}$ deformation alone (path P1). 
Therefore, except on those situations where the bottom of the valley 
runs parallel to either $\beta_{2}$ or $\beta_{4}$ axis the quadrupole 
and hexadecapole degrees of freedom cannot be decoupled  and the full 
fledged two dimensional GCM  has to be considered. In contrast, 
$\beta_{3}$ usually decouple from $\beta_{2}$ when the two degrees of 
freedom are considered together \cite{RodriguezGuzman2012}. As 
discussed below, an alternative to the 2D calculation could be the use of 
collective variables along A and B directions. 

For the valley in the oblate side one has $\beta_{4} \approx -0.37 
\beta_{2}$ and an energy exceeding 4 MeV the one of the prolate side. 
This energy difference between both minima implies that the oblate 
minimum is not playing an active role except for high lying excited 
states.

\begin{figure}
\includegraphics[angle=0,width=0.91\columnwidth]{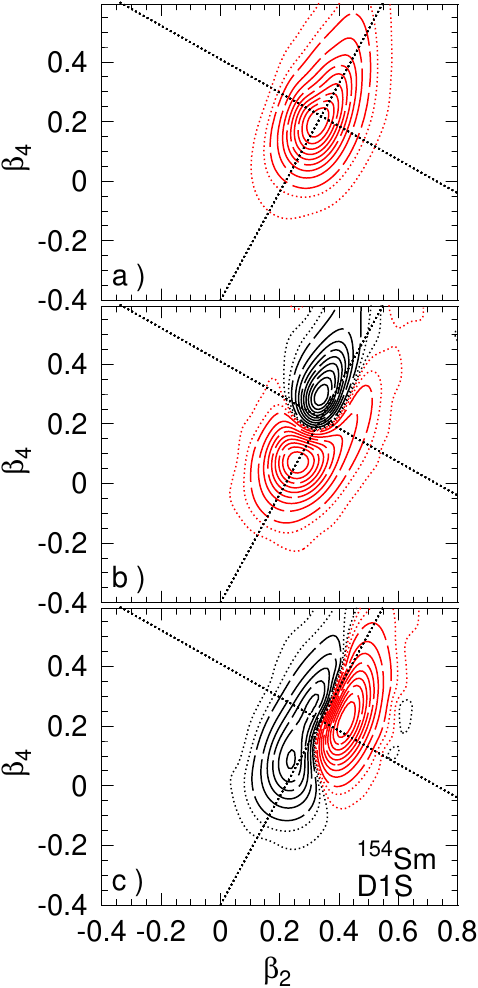}  
\caption{(Color online) Contour plot of the collective amplitudes 
$g_{\sigma} (\beta_{2},\beta_{4})$ as a function
of $\beta_{2}$ (horizontal axis) and $\beta_{4}$ (vertical axis) for the 
three lowest stated obtained in the 2D GCM. Black (red) contours correspond
to positive (negative) values of $g_{\sigma}$. Contours (dashed) correspond to
90 \%, 80 \%, $\ldots$ of the maximum value of $g_{\sigma}$ except the last two (dotted) that are
drawn at 5\% and 1\% of the maximum value. The perpendicular straight lines
are the same drawn in the HFB energy case in Fig. \ref{154SmEB2B4}.
}
\label{fig:154SmG} 
\end{figure}

The results of the calculation just 
considering $\beta_{2}$ as collective coordinate (path P1 in Fig. \ref{154SmEB2B4}) are shown in 
the three panels of Fig. \ref{154Sm1D_P1}. In Fig. \ref{154Sm1D_P1}(a) 
the HFB energy as a function of $\beta_{2}$ is shown. Two minima, one 
prolate and the other oblate are found, with the deeper prolate one the 
ground state. The oblate minimum has little 
influence in this case as it lies high up in energy as compared to the 
ground state. The $\beta_{4}$ deformation parameter is shown in Fig. 
\ref{154Sm1D_P1}(b). The deformation parameter decreases almost linearly 
for negative $\beta_{2}$ reaching a value close to zero at 
$\beta_{2}=0$. From there on, a linear increase is observed. The ground 
state at $\beta_{4}=0.21$ is rather large as compared to other 
regions of the nuclear chart. In  Fig. \ref{154Sm1D_P1}(c) the collective 
wave function $g_{\sigma} (\beta_{2})$ obtained in the 1D GCM is shown for the three lowest 
states. The wave functions are situated with respect to the $y$ axis 
according to the corresponding excitation energy of the collective 
state. Typical shapes, similar to the ones of the lowest states of the 
1D harmonic oscillator (HO), are seen for the collective wave 
functions. The ground state has a Gaussian distribution peaked at the 
ground-state $\beta_{2}$ deformation. The fist excited state has 
a node at the $\beta_{2}$ deformation of the ground state and decays like a Gaussian away from 
the minimum. The next excited state shows two 
nodes. All the wave functions show some distortions 
with respect to the 
ones of the HO. This can be attributed to deviations of collective
potential and inertia from the harmonic form
\footnote{By 
using the Gaussian Overlap approximation \cite{Ring1980}, the one 
dimensional GCM method can be approximated by a collective Schrodinger 
equation with a collective inertia given in terms of second derivatives 
of the Hamiltonian overlaps, see \cite{Ring1980} for details}. The 
absolute energy of the three states is plotted in Fig. 
\ref{154Sm1D_P1}(a) as bullets placed in the $\beta_{2}$ axes according to 
the average $\beta_{2}$ value of the correlated state. In the present case, 
the average $\beta_{2}$ value is rather similar for the three states. 
A similar calculation but using the $\beta_{4}$ deformation as 
collective coordinate (path P2 inf Fig. \ref{154SmEB2B4}) shows similar results as will be discussed below. 
This is not surprising if one compares the paths explored in both 
calculations and depicted in Fig. \ref{154SmEB2B4}. As expected 
\cite{Ring1980} (Chapter 7) the paths do not coincide with the bottom 
of the 2D valley due to the fact that in the 1D calculations the 
minimum of the energy is obtained subject to the corresponding 
constraint, i.e. along vertical (horizontal) lines in the $\beta_{2}$ 
($\beta_{4}$) potential energy surface (which is the quantity shown in 
Fig. \ref{154SmEB2B4}). However, the two paths are rather close to each 
other in the region close to the minimum and, as a consequence, the 
dynamics is rather similar in the two 1D GCM calculations.

\begin{table}
\caption{\label{tab:154Sm} Results of GCM calculations for $^{154}$Sm. 
First column labels the type of calculation, either a 1D ($\beta_{2}$ 
along path P1 or $\beta_{4}$ along path P2) or 2D 
($\beta_{2}-\beta_{4}$). The upper (last) three rows 
correspond to the results obtained with D1S (D1M*) parametrization of 
the Gogny force. 
The remaining columns are divided into sets of three. The first set shows
the correlation energy and the two moments $\beta_{2}$ 
and $\beta_{4}$ of the density distribution. The
other two sets show excitation energies and corresponding moments.
}
 \begin{center}
 \begin{ruledtabular}
  \begin{tabular}{c|ccc|ccc|ccc}
    	                    & $E_{c}$ & $\beta_{2}$ & $\beta_{4}$ & $E_{1}$ &  $\beta_{2}$ & $\beta_{4}$ &  $E_{2}$ &  $\beta_{2}$ & $\beta_{4}$ \\ \hline
$\beta_{2}$ (P1)            & 0.694   & 0.33        & 0.19        & 2.576   &  0.27        & 0.13        &  3.827   &  0.37        & 0.21   \\
$\beta_{4}$ (P2)            & 0.663   & 0.33        & 0.21        & 2.407   &  0.30        & 0.16        &  4.230   &  0.28        & 0.11   \\
$\beta_{2}-\beta_{4}$       & 1.239   & 0.33        & 0.21        & 2.635   &  0.30        & 0.17        &  3.059   &  0.37        & 0.20   \\ \hline
$\beta_{2}$ (P1)            & 0.637   & 0.32        & 0.18        & 2.933   &  0.29        & 0.14        &  4.261   & -0.23        & 0.10   \\
$\beta_{4}$ (P2)            & 0.707   & 0.32        & 0.19        & 2.447   &  0.28        & 0.12        &  4.615   &  0.28        & 0.12   \\
$\beta_{2}-\beta_{4}$       & 1.305   & 0.32        & 0.19        & 2.685   &  0.27        & 0.13        &  3.210   &  0.37        & 0.22      
  \end{tabular}
 \end{ruledtabular}
 \end{center}
\end{table}

At this point it is worth to discuss the results of another 1D calculation,
this time along the line marked as B in Fig.~\ref{154SmEB2B4} and perpendicular
to the bottom of the energy valley. In order to carry out the calculation
a set of HFB states was generated with $\beta_{2}$ in the range $[0.1,0.6]$
and $\beta_{4}$ constrained to be in the $\beta_{4}=-0.56 \beta_{2} + 0.41$ 
line. The results obtained are summarized in Fig. \ref{154Sm1D_B}. In panel
(a) the HFB energy shows a well defined and deep quadratic well. In panel (b)
the $\beta_{4}$ deformation follows a straight line as it should be. Finally,
in panel (c) the collective wave functions are shown. They follow closely
the expectations for a pure harmonic oscillator. The correlation energy
gained by the ground state in this 1D GCM calculation is 0.555 MeV. The
excitation energy of the lowest excited state is 3.075 MeV and at
roughly twice the excitation energy, 6.509 MeV, the second phonon state is 
located.  

Coming back to the two dimensional (2D) calculation, it is interesting 
to analyze the behavior of the collective wave function $g_{\sigma} 
(\beta_{2}, \beta_{4})$ solution to the Hill-Wheeler-Griffin equation. 
The quantities, corresponding to the three lowest solutions are shown 
in Fig. \ref{fig:154SmG}. The upper panel (a) corresponds to the ground 
state and shows the typical 2D Gaussian shape but tilted with respect 
to the $\beta_{2}$ and $\beta_{4}$ axes and closely aligned with 
respect to the A and B directions, shown as perpendicular dotted lines 
in the plot. The A and B directions run along the bottom of the energy 
valley (A) and the perpendicular direction (B). The middle panel (b) is 
for the first excited state. The shape corresponds to a Gaussian along 
the B principal axis. Along the direction of the principal axis A the 
shape of the wave function corresponds to a one-phonon state in a 1D 
HO. By comparing with the collective wave functions of 
Fig.~\ref{154Sm1D_P1} we conclude that this state corresponds to a 
collective  phonon in which the quadrupole and hexadecapole degrees of 
freedom are mixed together. Finally, the last panel (c) corresponds to 
the second excited state that can be interpreted as a 1D phonon along 
the B direction. We conclude from the present analysis that both 
quadrupole and hexadecapole degrees of freedom are strongly interleaved 
and it is better to talk about the A and B directions (or degrees of 
freedom) instead. It is interesting to note that A and B are given by 
linear relations (see above) in terms of $\beta_{2}$ and $\beta_{4}$ at 
least locally around the HFB minimum.

\begin{figure}
\includegraphics[angle=0,width=0.91\columnwidth]{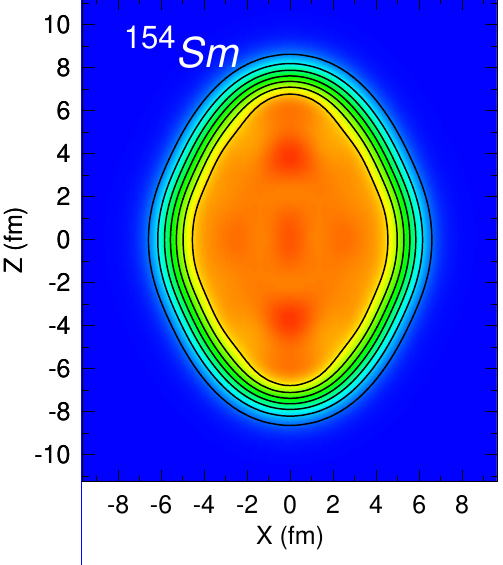}  
\caption{(Color online) The matter density distribution of the mean 
field ground state of $^{154}$Sm is shown. The mean field deformation 
parameters (Eq. \ref{betal}) for this state are $\beta_{2}=0.32$ and 
$\beta_{4}=0.21$
}
\label{fig:density} 
\end{figure}

In Table \ref{tab:154Sm} we show several quantities obtained in the GCM 
calculation in the 2D and 1D cases with both D1S and D1M* 
parametrizations of the Gogny force. The correlation energy $E_{c}$ gained 
in the two 1D calculations for D1S are similar corroborating the 
conclusion previously drawn about the equivalence of the 1D GCM results 
irrespective of the use of the $\beta_{2}$ and $\beta_{4}$ collective 
coordinates (paths P1 and P2). The 2D correlation energy is, accidentally, twice as large 
as the 1D one with a significant energy gain of 0.6 MeV due to the 
inclusion of the hexadecapole degree of freedom. This quantity is not 
negligible and its evolution with proton and neutron number could have 
significant impact on the reproduction of experimental binding 
energies. In this respect, modern energy density functionals (EDFs) are 
able to reach a root mean square (rms) deviation for the binding 
energies of around 700  keV \cite{Goriely2009,Ryssens2022}. For a 
systematic study of octupole correlation energies the reader is 
referred to Ref.~\cite{Robledo2015}. It is also important to note that the 2D correlation energy is consistently
given as the sum of the quantity obtained in the 1D calculation along path P1 plus the correlation
energy obtained along path B.
In this example as well as in the
other nuclei considered below the additional correlation energy gained in 
going from the 1D to the 2D case is similar to the one of the quadrupole
dynamics alone indicating a very slow convergence of the correlation energy with
the (even) multipole degrees considered in the GCM. Whether this is a 
feature of this specific region or a general trend should be analyzed by
extending this type of calculation to a significantly wider sample of nuclei in
the nuclear chart. It is to be expected that correlation energy corresponding
to multipoles of order six or higher should be significantly smaller than
the one of lower multipole orders and the general argument was given in
the introduction. This is, however, a still to be answered question that
deserves further consideration. 

The deformation parameters of the 
ground and  first excited states do not change much in going from the 
1D to the 2D GCM results. It is also remarkable that the ground state 
$\beta_{2}$ and $\beta_{4}$ GCM values are similar to the ones obtained 
at the HFB level. This result is not surprising as the ground state 
collective wave function is centered at the position of the HFB 
minimum.  However, the deformation parameters change significantly for 
the second excited states with respect to the ground state values. 
Regarding the excitation energies, the first excited state behaves 
similarly in the 1D and 2D calculations, but this is not the case for 
the second state. It is easy to understand the origin of the difference 
by looking at Fig. \ref{fig:154SmG}: the second excited state is a 1D 
phonon along path B, not present by definition in the 1D case. It is worth
to remember that this state is the first excited state in the 1D calculation
along path B discussed previously. Its 
excitation energy is slightly above 3 MeV and therefore could be one of 
the many $0^{+}$ excited states found in many nuclei in the region. 
However, the value of the excitation energy is perhaps a bit too high 
and therefore its features would be difficult to characterize 
experimentally. 

As a side comment, it is remarkable the mild dependence 
of the results with the parametrization of the Gogny force used. Both
share the same functional form, but the parameters were adjusted with
rather different targets in mind. For instance, D1M$^{*}$ produces
much better quality binding energies than D1S and it is also expected
to behave better in the neutron rich sector. 

A  comparison with the experimental data \cite{bnl} for the lowest excited 
$0^{+}$ states reveals a discrepancy of more than a factor of two 
between theory and experiment being the experimental data smaller than 
the theoretical predictions. In the $^{154}$Sm nucleus there are a 
couple of known excited $0^{+}$ at excitation energies slightly above 1 
MeV. However, it is not clear whether those two states can be 
unambiguously identified with a genuine $\beta$ vibration as discussed 
in \cite{Garrett2001,Garrett2018}. In these references, it is argued 
that $\beta$ vibrations should lie higher in energy due to the kind of 
excitations involved and its excitation energy very sensitive to 
pairing effects, not taken explicitly into account in the present 
description. On the other hand, the theoretical description includes a 
limited set of collective degrees of freedom and it is very likely that 
triaxial and pairing effects can play a role in the properties of the 
first excited state. One also should not forget that the GCM formalism 
do not take into account collective momentum degrees of freedom. The 
impact of those on the dynamics is not well studied but based on the 
large differences between collective inertias for fission obtained with 
the Adiabatic Time Dependent (ATD) and the GCM frameworks 
\cite{Giuliani2018} an important reduction of the excitation energies 
consequence of the use of collective momentum degrees of freedom is to 
be expected. There is additional insight pointing to this effect coming 
from  Random Phase Approximation results \cite{Lechaftois2015}. As a 
conclusion, all the above effects should be considered to have a more 
precise estimation of the hexadecapole $K= 0^{+}$ vibrational 
excitation energy. It is likely that its value will be lower than the 
present prediction and therefore more likely to be characterized 
experimentally.

The quadrupole deformation parameter is larger than the one in 
\cite{GOTZ19721} and also in \cite{Moeller2016} but the difference 
could be attributed to the definition of $\beta_{2}$. If one uses the 
definition of $\beta_{2}$ with $\langle r^{2} \rangle$ instead of $3/5 
R^{2}$ with $R=r_{0} A^{1/3}$ used here, smaller values are obtained 
(typically 20 \% smaller ). The same also holds true for $\beta_{4}$ 
but in this case the deviation of the present results with respect to 
the ones of Refs.~\cite{GOTZ19721,Moeller2016} is as large as a factor 
of two. A comparison with experimental data of Refs.
\cite{Erb1972,Ronningen1977} also indicates an overestimation of 
$\beta_{4}$ with respect to the experiment by a factor of 2.

As the deformation parameters for $^{154}$Sm and all the studied nuclei 
considered in this paper (see below) are rather large, larger than the 
ones predicted by Moller in \cite{Moeller2016}, it is therefore 
instructive to have a look at the spatial distribution of the matter 
density for the HFB solution at the minimum shown in 
Fig.~\ref{fig:density}. One observes the typical diamond like shape 
characteristic of positive $\beta_{4}$ values.

\subsection{Potential energy surfaces of Gd and Sm \label{sec:pesGdSm}}

\begin{figure}
\includegraphics[angle=0,width=0.95\columnwidth]{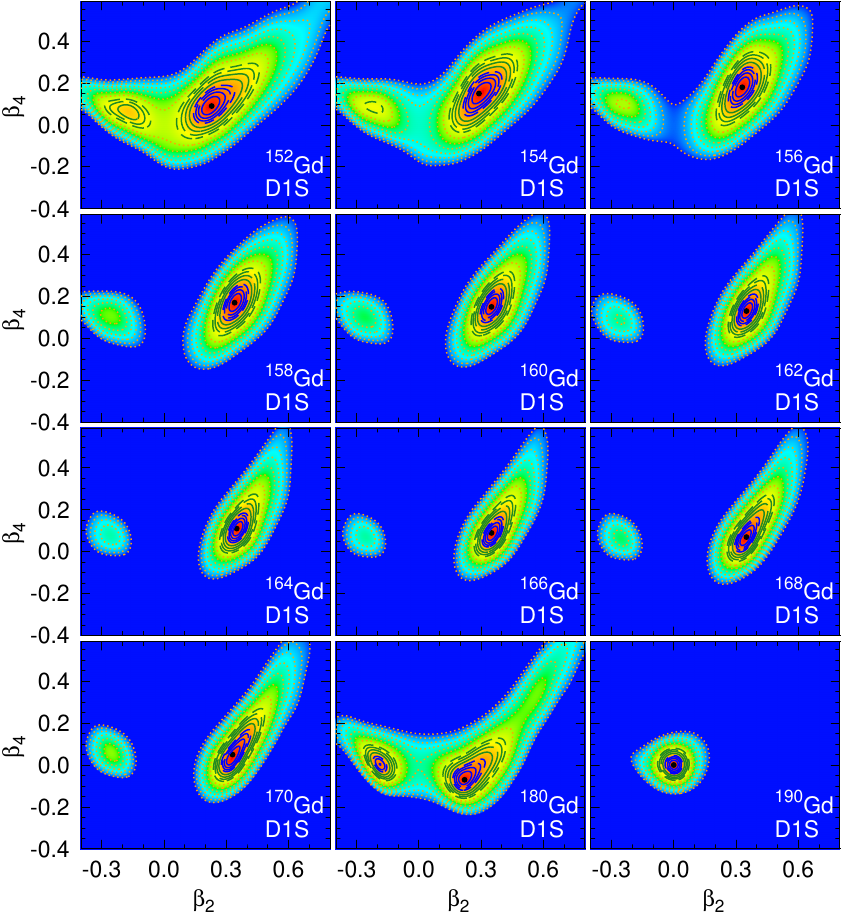}  
\caption{(Color online) Potential energy surfaces as a function of
$\beta_{2}$ and $\beta_{4}$ deformation parameters for the Gd isotopes 
considered. The results are obtained with Gogny D1S.
}
\label{fig:MFPES_Gd} 
\end{figure}

The HFB energy as a function of $\beta_{2}$ and $\beta_{4}$ for the 
nuclei in the isotopic chain of Gd is shown in Fig.~\ref{fig:MFPES_Gd}. 
Most of the energies show a valley whose bottom roughly follows a 
straight line with positive slope in the $\beta_{2}-\beta_{4}$ plane 
for prolate deformations. The valley bends at $\beta_{2}\approx 0$ to 
acquire a negative slope but the excitation energy in that region is 
large and its effect on the ground state low energy dynamic can be 
disregarded. The only two exceptions are the isotope of $^{180}$Gd 
where the oblate minimum lies quite low in energy and the isotope of 
$^{190}$Gd with magic neutron number $N=126$ which shows the 
characteristic behavior of a spherical nucleus. The ground state 
minimum takes place at rather large $\beta_{4}$ values and all of them 
are  prolate deformed. For the Sm isotopes to be discussed later on, 
the results look very similar and therefore are not shown here. The HFB 
energy looks rather similar to the $^{154}$Sm one discussed in the 
previous section and therefore all the considerations there apply to 
all the nuclei in the chain exception made of $^{180}$Gd where 
prolate and oblate minima coexist and $^{190}$Gd which is a semi-magic spherical
nucleus.

\subsection{Generator coordinate method results for Gd and Sm \label{sec:gcmGdSm}}

\begin{figure}
\includegraphics[angle=0,width=0.95\columnwidth]{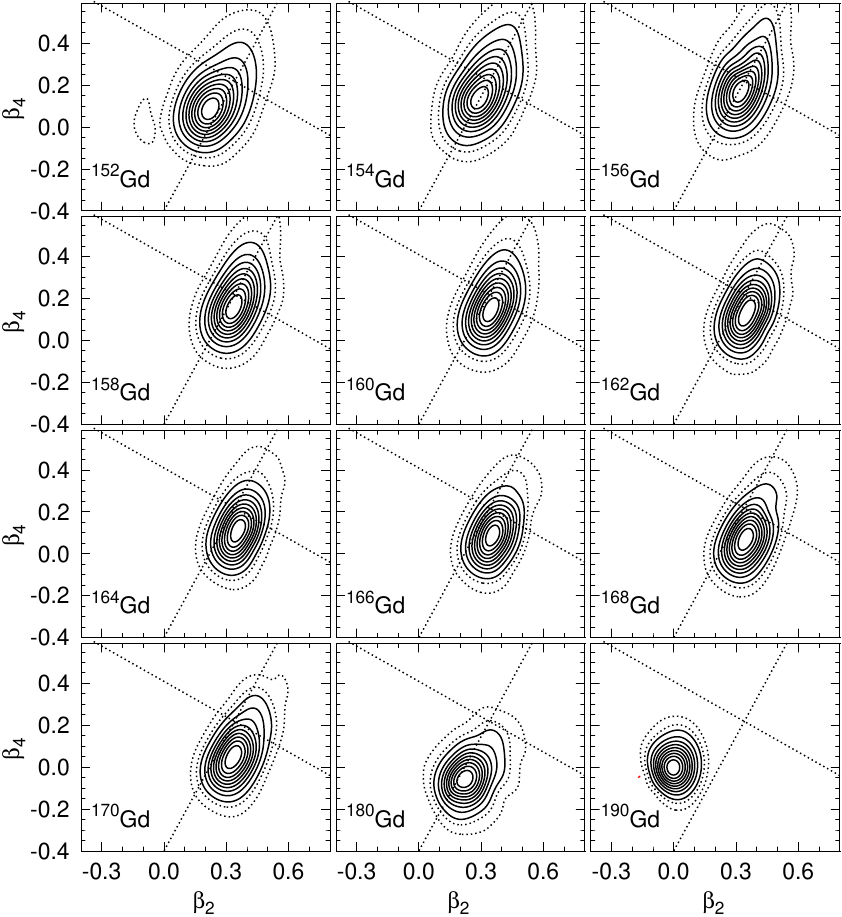}  
\caption{(Color online) Contour plots of the collective ground state 
wave function $g_{0}(\beta_{2},\beta_{4})$ of Eq. (\ref{cll-wfs-HW}) 
obtained by solving the GHW equation in the Gd isotopes. The results 
are obtained with Gogny D1S. }
\label{fig:G0_Gd} 
\end{figure}

In Fig. \ref{fig:G0_Gd} the collective amplitude corresponding to the 
ground state is shown for the considered nuclei. Following the 
discussion of the $^{154}$Sm case, one clearly identify the 
characteristic two dimensional Gaussian shape with principal axes 
aligned roughly in the same directions A and B s in the $^{154}$Sm case.

\begin{figure}
\includegraphics[angle=0,width=0.95\columnwidth]{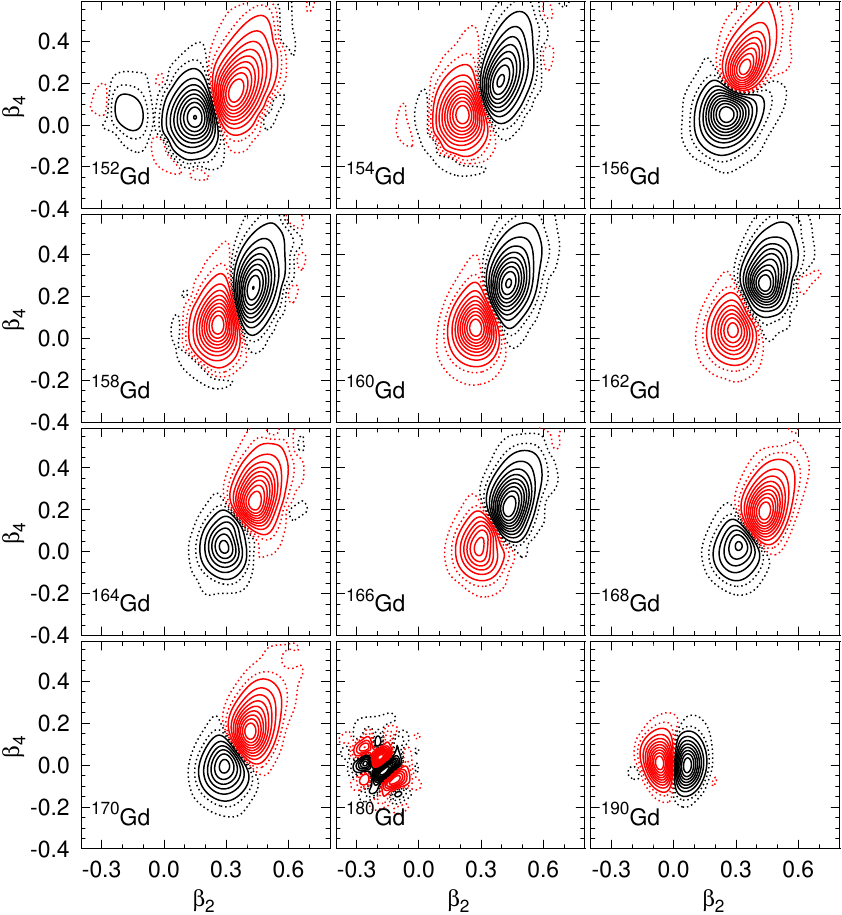}  
\caption{(Color online) Contour plots of the collective first excited 
state wave function $g_{1}(\beta_{2},\beta_{4})$ of Eq. (\ref{cll-wfs-HW}) 
obtained by solving the GHW equation in the Gd isotopes. Black (red) 
contour lines correspond to positive (negative) values of the 
collective wave function $g_{1}(\beta_{2},\beta_{4})$. The results are 
obtained with Gogny D1S.
}
\label{fig:G1_Gd} 
\end{figure}

In Fig. \ref{fig:G1_Gd} the collective amplitude corresponding to the 
first excited state is shown for the considered nuclei. Following the 
discussion of the $^{154}$Sm case, one clearly identify the 
characteristic two dimensional shape corresponding to a 1D phonon in 
the collective variable along the bottom of the valley (path A). In the 
$^{180}$Gd case, the first excited state corresponds to a shape 
coexisting oblate configuration and the collective amplitude is 
concentrated in the oblate minimum. In the $^{190}$Gd magic nucleus, 
the first excited state is a pure $\beta_{2}$ vibration. 

In Table \ref{tab:Gd} the results obtained for the ground state and two 
first excited states in the 2D $\beta_{2}-\beta_{4}$ calculations are 
given as a function of mass number $A$ for the Gd isotopes. In the 
second column the correlation energy gained by the collective motion 
$\beta_{2}-\beta_{4}$ on top of the mean field ground state 
energy is given. Overall, one observes a gain of around 1.2 MeV 
correlation energy, but the behavior as a function of $A$ is not 
constant with a minimum of 0.99 MeV for $A=164$ and a maximum of 1.46 
MeV for $A=154$. The additional 600 keV binding energy gain associated
with the hexadecapole degree of freedom can be important for a proper 
description of binding energies with the accuracy required by modern 
applications \cite{Goriely2009,Ryssens2022}. In the third and fourth 
columns the $\beta_{2}$ and $\beta_{4}$ deformation parameters are 
given, Our $\beta_{2}$ parameters are typically 25\% larger than the 
ones given by Moller \cite{Moeller2016} and the only isotope where they 
agree is the spherical $^{190}$Gd. On the other hand, our $\beta_{4}$ 
values are a factor of two larger. The GCM ground state deformation 
parameters are similar to the ones obtained at the mean field level. In 
the next three columns, the excitation energy of the first excited 
state $0^{+}$ (phonon along the A direction) along with its $\beta_{2}$ and $\beta_{4}$ deformation 
parameters are given. The excitation energy ranges from 0.6 to 3.7 MeV 
depending on the isotope and both deformation parameters are slightly 
larger than those of the ground state. In $^{180}$Gd, the first excited 
$0^{+}$ is oblate and lies at a quite low excitation energy of 612 keV 
with a $\beta_{2}=-0.19$ and zero hexadecapole deformation. The Gd 
isotopes with A=152 and 154  show prolate-oblate shape coexistence 
that manifest in a different structure of the collective wave function 
of the first excited state (see Fig. \ref{fig:G1_Gd}). As a consequence, 
the excitation energy of the first excited state is relatively low with 
values of 1.5 and 1.9 MeV, respectively. The same holds true for the 
isotopes with A=168 and 170 but with a less pronounced prolate-oblate 
mixing. For the intermediate isotopes, without shape coexistence the 
energy goes up to around 3.5 MeV. For the second excited $0^{+}$ the 
excitation energies follow the same pattern as for the first excited 
state associated to the existence of prolate-oblate shape coexistence. 
For mass numbers around 162 it goes up to around 4.8 MeV. 
Interestingly, the $\beta_{2}$ deformation of the second excited state 
becomes negative for A=152, 154 and 170 as a clear manifestation of 
prolate-oblate mixing. All the remaining isotopes have  similar 
deformation parameters as the ground state. The excitation energy of 
the second excited state (phonon along B direction) appears too high except in those cases where 
prolate-oblate shape coexistence is present. As discussed in the 
previous subsection in the $^{154}$Sm case, some missing degrees of 
freedom might reduce the excitation energy a bit. Using the typical 
reduction of a factor 0.7 consequence of  considering ATD versus GCM 
inertias (a simple way to take into account momentum-like collective 
coordinates) one could expect excitation energies for the phonon along B 
direction to come down to 3-3.5 MeV which is perhaps too high to be 
characterized experimentally. For those cases where prolate-oblate 
shape coexistence is present the above reduction factor will bring the 
excitation energy to a quite low value but the price to pay would be to 
disentangle the impact of shape coexistence in the characteristics of 
the vibrational state. 


\begin{table}
	\caption{\label{tab:Gd} Same as Table \ref{tab:154Sm} but for the 2D GCM
	calculation of all the Gd isotopes considered and using the D1S parametrization
	of the force. 
	}
 \begin{center}
 \begin{ruledtabular}
  \begin{tabular}{c|ccc|ccc|ccc}
   A  & $E_{c}$ & $\beta_{2}$ & $\beta_{4}$  & $E_{1}$  & $\beta_{2}$ & $\beta_{4}$ &  $E_{2}$  &  $\beta_{2}$ & $\beta_{4}$ \\ \hline
  152 & 1.384   &  0.23       &  0.10        & 1.547    &  0.27       &  0.13       &  1.638    & -0.16        &    0.06     \\
  154 & 1.464   &  0.30       &  0.16        & 1.874    &  0.29       &  0.13       &  2.790    & -0.21        &    0.08     \\
  156 & 1.325   &  0.33       &  0.18        & 2.631    &  0.29       &  0.14       &  2.836    &  0.38        &    0.20     \\
  158 & 1.290   &  0.34       &  0.17        & 3.191    &  0.35       &  0.16       &  3.876    &  0.35        &    0.15     \\
  160 & 1.148   &  0.35       &  0.15        & 3.502    &  0.36       &  0.16       &  4.422    &  0.38        &    0.14     \\
  162 & 1.153   &  0.35       &  0.13        & 3.710    &  0.39       &  0.20       &  4.836    &  0.38        &    0.14     \\
  164 & 0.991   &  0.36       &  0.11        & 3.665    &  0.41       &  0.21       &  4.830    &  0.35        &    0.11     \\
  166 & 1.028   &  0.36       &  0.09        & 3.114    &  0.41       &  0.18       &  3.854    &  0.37        &    0.10     \\
  168 & 1.141   &  0.35       &  0.07        & 2.217    &  0.42       &  0.17       &  4.514    &  0.36        &    0.09     \\
  170 & 1.233   &  0.34       &  0.05        & 1.887    &  0.39       &  0.12       &  4.161    & -0.25        &    0.06     \\
  180 & 1.491   &  0.23       & -0.05        & 0.612    & -0.19       &  0.00       &  1.776    &  0.30        &    0.01     \\
  190 & 1.196   &  0.00       &  0.00        & 3.987    & -0.00       &  0.00       &  5.735    & -0.00        &    0.01     
  \end{tabular}
 \end{ruledtabular}
 \end{center}
\end{table}

In Table \ref{tab:Sm} the results obtained for the Sm isotopic chain are
presented. The features of the ground, first and second excited states are
very similar to the ones of the corresponding Gd isotopes with the same
mass number plus two. It becomes apparent that a change of two units in 
proton number is not changing in a relevant way the Gd results except in
very specific situations like the $\beta_{2}$ deformation of the second
excited state in $^{152}$Sm. These specific cases can be traced back to
a subtle interplay of the collective wave functions associated to prolate-oblate
shape coexistence in those systems.

\begin{table}
	\caption{\label{tab:Sm} Same as Table \ref{tab:154Sm} but for the 2D GCM
	calculation of all the Sm isotopes considered and using the D1S parametrization
	of the force. 
	}
 \begin{center}
 \begin{ruledtabular}
  \begin{tabular}{c|ccc|ccc|ccc}
   A  & $E_{c}$  & $\beta_{2}$ & $\beta_{4}$  & $E_{1}$   & $\beta_{2}$ & $\beta_{4}$ &  $E_{2}$  &  $\beta_{2}$ & $\beta_{4}$ \\ \hline
  150 &  1.338   &   0.23      &   0.10       &   1.496   &   0.28      &   0.14      &   1.996   &  -0.16       &   0.06      \\
  152 &  1.403   &   0.31      &   0.18       &   1.879   &   0.28      &   0.12      &   3.076   &   0.31       &   0.17      \\
  154 &  1.239   &   0.33      &   0.21       &   2.635   &   0.30      &   0.17      &   3.059   &   0.37       &   0.20      \\
  156 &  1.174   &   0.35      &   0.20       &   3.319   &   0.33      &   0.15      &   4.083   &   0.36       &   0.17      \\
  158 &  1.072   &   0.35      &   0.18       &   3.464   &   0.34      &   0.16      &   4.629   &   0.39       &   0.16      \\
  160 &  0.968   &   0.36      &   0.15       &   3.554   &   0.37      &   0.19      &   4.943   &   0.40       &   0.16      \\
  162 &  0.961   &   0.36      &   0.13       &   3.390   &   0.39      &   0.20      &   4.845   &   0.36       &   0.14      \\
  164 &  1.046   &   0.36      &   0.11       &   2.725   &   0.40      &   0.18      &   4.233   &   0.36       &   0.11      \\
  166 &  1.194   &   0.36      &   0.10       &   1.648   &   0.41      &   0.16      &   4.172   &   0.35       &   0.09      \\
  168 &  1.371   &   0.35      &   0.09       &   1.452   &   0.37      &   0.11      &   3.951   &  -0.26       &   0.07      
  \end{tabular}
 \end{ruledtabular}
 \end{center}
\end{table}

\section{Summary\label{sec:summary}}

Systematic HFB calculations with the Gogny D1S force show several 
regions of the nuclear chart where the ground state hexadecapole 
deformation is non zero. In this paper the region corresponding with 
$Z=62$ and 64 (Sm and Gd) and positive $\beta_{4}$ values is studied 
with the GCM method for the $K=0^{+}$ quadrupole and hexadecapole 
degrees of freedom. The gain in binding energy due to those 
correlations is computed as well as the position of the first and 
second $0^{+}$ states corresponding to  vibrational states along 
collective degrees of freedom where quadrupole and hexadecapole are 
strongly interleaved. For some of the isotopes considered 
prolate-oblate shape coexistence impacts excitation energies and 
deformation parameters of excited  states in a substantial way. For the 
more pure vibrational states showing up in some other isotopes 
excitation energies come up too high to be amenable to an easy 
experimental characterization. A discussion of relevant missing degrees 
of freedom that could reduce the excitation energies is presented. We 
conclude that the physics brought by considering the hexadecapole 
degree of freedom is not trivial and its study is worth further 
consideration. In a forthcoming publication we plan to extend the 
present analysis to nuclei in the rare earth region with negative 
$\beta_{4}$ values in their ground states.

\begin{acknowledgments}
The work of LMR was supported by Spanish Agencia Estatal de 
Investigacion (AEI) of the Ministry of Science and Innovation under 
Grant No. PID2021-127890NB-I00. C.V.N.K. acknowledges the Erasmus Mundus 
Master on Nuclear Physics (Grant agreement number 2019-2130) supported 
by the Erasmus+ Programme of the European Union for a scholarship.
\end{acknowledgments}


\begin{thebibliography}{38}%
\makeatletter
\providecommand \@ifxundefined [1]{%
 \@ifx{#1\undefined}
}%
\providecommand \@ifnum [1]{%
 \ifnum #1\expandafter \@firstoftwo
 \else \expandafter \@secondoftwo
 \fi
}%
\providecommand \@ifx [1]{%
 \ifx #1\expandafter \@firstoftwo
 \else \expandafter \@secondoftwo
 \fi
}%
\providecommand \natexlab [1]{#1}%
\providecommand \enquote  [1]{``#1''}%
\providecommand \bibnamefont  [1]{#1}%
\providecommand \bibfnamefont [1]{#1}%
\providecommand \citenamefont [1]{#1}%
\providecommand \href@noop [0]{\@secondoftwo}%
\providecommand \href [0]{\begingroup \@sanitize@url \@href}%
\providecommand \@href[1]{\@@startlink{#1}\@@href}%
\providecommand \@@href[1]{\endgroup#1\@@endlink}%
\providecommand \@sanitize@url [0]{\catcode `\\12\catcode `\$12\catcode
  `\&12\catcode `\#12\catcode `\^12\catcode `\_12\catcode `\%12\relax}%
\providecommand \@@startlink[1]{}%
\providecommand \@@endlink[0]{}%
\providecommand \url  [0]{\begingroup\@sanitize@url \@url }%
\providecommand \@url [1]{\endgroup\@href {#1}{\urlprefix }}%
\providecommand \urlprefix  [0]{URL }%
\providecommand \Eprint [0]{\href }%
\providecommand \doibase [0]{https://doi.org/}%
\providecommand \selectlanguage [0]{\@gobble}%
\providecommand \bibinfo  [0]{\@secondoftwo}%
\providecommand \bibfield  [0]{\@secondoftwo}%
\providecommand \translation [1]{[#1]}%
\providecommand \BibitemOpen [0]{}%
\providecommand \bibitemStop [0]{}%
\providecommand \bibitemNoStop [0]{.\EOS\space}%
\providecommand \EOS [0]{\spacefactor3000\relax}%
\providecommand \BibitemShut  [1]{\csname bibitem#1\endcsname}%
\let\auto@bib@innerbib\@empty
\bibitem [{\citenamefont {Hilaire}\ and\ \citenamefont
  {Girod}(2005)}]{Hilaire2005}%
  \BibitemOpen
  \bibfield  {author} {\bibinfo {author} {\bibfnamefont {S.}~\bibnamefont
  {Hilaire}}\ and\ \bibinfo {author} {\bibfnamefont {M.}~\bibnamefont
  {Girod}},\ }\href {https://doi.org/10.1140/epja/i2007-10450-2} {\bibfield
  {journal} {\bibinfo  {journal} {The European Physical Journal A - Hadrons and
  Nuclei}\ }\textbf {\bibinfo {volume} {33}},\ \bibinfo {pages} {237} (\bibinfo
  {year} {2005})}\BibitemShut {NoStop}%
\bibitem [{\citenamefont {Möller}\ \emph {et~al.}(2016)\citenamefont
  {Möller}, \citenamefont {Sierk}, \citenamefont {Ichikawa},\ and\
  \citenamefont {Sagawa}}]{Moeller2016}%
  \BibitemOpen
  \bibfield  {author} {\bibinfo {author} {\bibfnamefont {P.}~\bibnamefont
  {Möller}}, \bibinfo {author} {\bibfnamefont {A.}~\bibnamefont {Sierk}},
  \bibinfo {author} {\bibfnamefont {T.}~\bibnamefont {Ichikawa}},\ and\
  \bibinfo {author} {\bibfnamefont {H.}~\bibnamefont {Sagawa}},\ }\href
  {https://doi.org/https://doi.org/10.1016/j.adt.2015.10.002} {\bibfield
  {journal} {\bibinfo  {journal} {Atomic Data and Nuclear Data Tables}\
  }\textbf {\bibinfo {volume} {109-110}},\ \bibinfo {pages} {1} (\bibinfo
  {year} {2016})}\BibitemShut {NoStop}%
\bibitem [{\citenamefont {{Scamps, Guillaume}}\ \emph
  {et~al.}(2021)\citenamefont {{Scamps, Guillaume}}, \citenamefont {{Goriely,
  Stephane}}, \citenamefont {{Olsen, Erik}}, \citenamefont {{Bender,
  Michael}},\ and\ \citenamefont {{Ryssens, Wouter}}}]{Scamps2021}%
  \BibitemOpen
  \bibfield  {author} {\bibinfo {author} {\bibnamefont {{Scamps, Guillaume}}},
  \bibinfo {author} {\bibnamefont {{Goriely, Stephane}}}, \bibinfo {author}
  {\bibnamefont {{Olsen, Erik}}}, \bibinfo {author} {\bibnamefont {{Bender,
  Michael}}},\ and\ \bibinfo {author} {\bibnamefont {{Ryssens, Wouter}}},\
  }\href {https://doi.org/10.1140/epja/s10050-021-00642-1} {\bibfield
  {journal} {\bibinfo  {journal} {Eur. Phys. J. A}\ }\textbf {\bibinfo {volume}
  {57}},\ \bibinfo {pages} {333} (\bibinfo {year} {2021})}\BibitemShut
  {NoStop}%
\bibitem [{\citenamefont {Lalazissis}\ \emph {et~al.}(1999)\citenamefont
  {Lalazissis}, \citenamefont {Raman},\ and\ \citenamefont
  {Ring}}]{Lalazissis1999}%
  \BibitemOpen
  \bibfield  {author} {\bibinfo {author} {\bibfnamefont {G.}~\bibnamefont
  {Lalazissis}}, \bibinfo {author} {\bibfnamefont {S.}~\bibnamefont {Raman}},\
  and\ \bibinfo {author} {\bibfnamefont {P.}~\bibnamefont {Ring}},\ }\href
  {https://doi.org/https://doi.org/10.1006/adnd.1998.0795} {\bibfield
  {journal} {\bibinfo  {journal} {Atomic Data and Nuclear Data Tables}\
  }\textbf {\bibinfo {volume} {71}},\ \bibinfo {pages} {1} (\bibinfo {year}
  {1999})}\BibitemShut {NoStop}%
\bibitem [{\citenamefont {Garrett}\ \emph {et~al.}(2005)\citenamefont
  {Garrett}, \citenamefont {Kulp}, \citenamefont {Wood}, \citenamefont
  {Bandyopadhyay}, \citenamefont {Christen}, \citenamefont {Choudry},
  \citenamefont {Dewald}, \citenamefont {Fitzler}, \citenamefont {Fransen},
  \citenamefont {Jessen}, \citenamefont {Jolie}, \citenamefont {Kloezer},
  \citenamefont {Kudejova}, \citenamefont {Kumar}, \citenamefont {Lesher},
  \citenamefont {Linnemann}, \citenamefont {Lisetskiy}, \citenamefont {Martin},
  \citenamefont {Masur}, \citenamefont {McEllistrem}, \citenamefont {Möller},
  \citenamefont {Mynk}, \citenamefont {Orce}, \citenamefont {Pejovic},
  \citenamefont {Pissulla}, \citenamefont {Regis}, \citenamefont {Schiller},
  \citenamefont {Tonev},\ and\ \citenamefont {Yates}}]{Garrett2005}%
  \BibitemOpen
  \bibfield  {author} {\bibinfo {author} {\bibfnamefont {P.~E.}\ \bibnamefont
  {Garrett}}, \bibinfo {author} {\bibfnamefont {W.~D.}\ \bibnamefont {Kulp}},
  \bibinfo {author} {\bibfnamefont {J.~L.}\ \bibnamefont {Wood}}, \bibinfo
  {author} {\bibfnamefont {D.}~\bibnamefont {Bandyopadhyay}}, \bibinfo {author}
  {\bibfnamefont {S.}~\bibnamefont {Christen}}, \bibinfo {author}
  {\bibfnamefont {S.}~\bibnamefont {Choudry}}, \bibinfo {author} {\bibfnamefont
  {A.}~\bibnamefont {Dewald}}, \bibinfo {author} {\bibfnamefont
  {A.}~\bibnamefont {Fitzler}}, \bibinfo {author} {\bibfnamefont
  {C.}~\bibnamefont {Fransen}}, \bibinfo {author} {\bibfnamefont
  {K.}~\bibnamefont {Jessen}}, \bibinfo {author} {\bibfnamefont
  {J.}~\bibnamefont {Jolie}}, \bibinfo {author} {\bibfnamefont
  {A.}~\bibnamefont {Kloezer}}, \bibinfo {author} {\bibfnamefont
  {P.}~\bibnamefont {Kudejova}}, \bibinfo {author} {\bibfnamefont
  {A.}~\bibnamefont {Kumar}}, \bibinfo {author} {\bibfnamefont {S.~R.}\
  \bibnamefont {Lesher}}, \bibinfo {author} {\bibfnamefont {A.}~\bibnamefont
  {Linnemann}}, \bibinfo {author} {\bibfnamefont {A.}~\bibnamefont
  {Lisetskiy}}, \bibinfo {author} {\bibfnamefont {D.}~\bibnamefont {Martin}},
  \bibinfo {author} {\bibfnamefont {M.}~\bibnamefont {Masur}}, \bibinfo
  {author} {\bibfnamefont {M.~T.}\ \bibnamefont {McEllistrem}}, \bibinfo
  {author} {\bibfnamefont {O.}~\bibnamefont {Möller}}, \bibinfo {author}
  {\bibfnamefont {M.}~\bibnamefont {Mynk}}, \bibinfo {author} {\bibfnamefont
  {J.~N.}\ \bibnamefont {Orce}}, \bibinfo {author} {\bibfnamefont
  {P.}~\bibnamefont {Pejovic}}, \bibinfo {author} {\bibfnamefont
  {T.}~\bibnamefont {Pissulla}}, \bibinfo {author} {\bibfnamefont {J.~M.}\
  \bibnamefont {Regis}}, \bibinfo {author} {\bibfnamefont {A.}~\bibnamefont
  {Schiller}}, \bibinfo {author} {\bibfnamefont {D.}~\bibnamefont {Tonev}},\
  and\ \bibinfo {author} {\bibfnamefont {S.~W.}\ \bibnamefont {Yates}},\ }\href
  {https://doi.org/10.1088/0954-3899/31/10/087} {\bibfield  {journal} {\bibinfo
   {journal} {Journal of Physics G: Nuclear and Particle Physics}\ }\textbf
  {\bibinfo {volume} {31}},\ \bibinfo {pages} {S1855} (\bibinfo {year}
  {2005})}\BibitemShut {NoStop}%
\bibitem [{\citenamefont {Phillips}\ \emph {et~al.}(2010)\citenamefont
  {Phillips}, \citenamefont {Garrett}, \citenamefont {Lo~Iudice}, \citenamefont
  {Sushkov}, \citenamefont {Bettermann}, \citenamefont {Braun}, \citenamefont
  {Burke}, \citenamefont {Demand}, \citenamefont {Faestermann}, \citenamefont
  {Finlay}, \citenamefont {Green}, \citenamefont {Hertenberger}, \citenamefont
  {Leach}, \citenamefont {Kr\"ucken}, \citenamefont {Schumaker}, \citenamefont
  {Svensson}, \citenamefont {Wirth},\ and\ \citenamefont
  {Wong}}]{Phillips2010}%
  \BibitemOpen
  \bibfield  {author} {\bibinfo {author} {\bibfnamefont {A.~A.}\ \bibnamefont
  {Phillips}}, \bibinfo {author} {\bibfnamefont {P.~E.}\ \bibnamefont
  {Garrett}}, \bibinfo {author} {\bibfnamefont {N.}~\bibnamefont {Lo~Iudice}},
  \bibinfo {author} {\bibfnamefont {A.~V.}\ \bibnamefont {Sushkov}}, \bibinfo
  {author} {\bibfnamefont {L.}~\bibnamefont {Bettermann}}, \bibinfo {author}
  {\bibfnamefont {N.}~\bibnamefont {Braun}}, \bibinfo {author} {\bibfnamefont
  {D.~G.}\ \bibnamefont {Burke}}, \bibinfo {author} {\bibfnamefont {G.~A.}\
  \bibnamefont {Demand}}, \bibinfo {author} {\bibfnamefont {T.}~\bibnamefont
  {Faestermann}}, \bibinfo {author} {\bibfnamefont {P.}~\bibnamefont {Finlay}},
  \bibinfo {author} {\bibfnamefont {K.~L.}\ \bibnamefont {Green}}, \bibinfo
  {author} {\bibfnamefont {R.}~\bibnamefont {Hertenberger}}, \bibinfo {author}
  {\bibfnamefont {K.~G.}\ \bibnamefont {Leach}}, \bibinfo {author}
  {\bibfnamefont {R.}~\bibnamefont {Kr\"ucken}}, \bibinfo {author}
  {\bibfnamefont {M.~A.}\ \bibnamefont {Schumaker}}, \bibinfo {author}
  {\bibfnamefont {C.~E.}\ \bibnamefont {Svensson}}, \bibinfo {author}
  {\bibfnamefont {H.-F.}\ \bibnamefont {Wirth}},\ and\ \bibinfo {author}
  {\bibfnamefont {J.}~\bibnamefont {Wong}},\ }\href
  {https://doi.org/10.1103/PhysRevC.82.034321} {\bibfield  {journal} {\bibinfo
  {journal} {Phys. Rev. C}\ }\textbf {\bibinfo {volume} {82}},\ \bibinfo
  {pages} {034321} (\bibinfo {year} {2010})}\BibitemShut {NoStop}%
\bibitem [{\citenamefont {Hartley}\ \emph {et~al.}(2020)\citenamefont
  {Hartley}, \citenamefont {Kondev}, \citenamefont {Savard}, \citenamefont
  {Clark}, \citenamefont {Ayangeakaa}, \citenamefont {Bottoni}, \citenamefont
  {Carpenter}, \citenamefont {Copp}, \citenamefont {Hicks}, \citenamefont
  {Hoffman}, \citenamefont {Janssens}, \citenamefont {Lauritsen}, \citenamefont
  {Orford}, \citenamefont {Sethi},\ and\ \citenamefont {Zhu}}]{Hartley2020}%
  \BibitemOpen
  \bibfield  {author} {\bibinfo {author} {\bibfnamefont {D.~J.}\ \bibnamefont
  {Hartley}}, \bibinfo {author} {\bibfnamefont {F.~G.}\ \bibnamefont {Kondev}},
  \bibinfo {author} {\bibfnamefont {G.}~\bibnamefont {Savard}}, \bibinfo
  {author} {\bibfnamefont {J.~A.}\ \bibnamefont {Clark}}, \bibinfo {author}
  {\bibfnamefont {A.~D.}\ \bibnamefont {Ayangeakaa}}, \bibinfo {author}
  {\bibfnamefont {S.}~\bibnamefont {Bottoni}}, \bibinfo {author} {\bibfnamefont
  {M.~P.}\ \bibnamefont {Carpenter}}, \bibinfo {author} {\bibfnamefont
  {P.}~\bibnamefont {Copp}}, \bibinfo {author} {\bibfnamefont {K.}~\bibnamefont
  {Hicks}}, \bibinfo {author} {\bibfnamefont {C.~R.}\ \bibnamefont {Hoffman}},
  \bibinfo {author} {\bibfnamefont {R.~V.~F.}\ \bibnamefont {Janssens}},
  \bibinfo {author} {\bibfnamefont {T.}~\bibnamefont {Lauritsen}}, \bibinfo
  {author} {\bibfnamefont {R.}~\bibnamefont {Orford}}, \bibinfo {author}
  {\bibfnamefont {J.}~\bibnamefont {Sethi}},\ and\ \bibinfo {author}
  {\bibfnamefont {S.}~\bibnamefont {Zhu}},\ }\href
  {https://doi.org/10.1103/PhysRevC.101.044301} {\bibfield  {journal} {\bibinfo
   {journal} {Phys. Rev. C}\ }\textbf {\bibinfo {volume} {101}},\ \bibinfo
  {pages} {044301} (\bibinfo {year} {2020})}\BibitemShut {NoStop}%
\bibitem [{\citenamefont {Magierski}\ \emph {et~al.}(1995)\citenamefont
  {Magierski}, \citenamefont {Heenen},\ and\ \citenamefont
  {Nazarewicz}}]{Magierski1995}%
  \BibitemOpen
  \bibfield  {author} {\bibinfo {author} {\bibfnamefont {P.}~\bibnamefont
  {Magierski}}, \bibinfo {author} {\bibfnamefont {P.-H.}\ \bibnamefont
  {Heenen}},\ and\ \bibinfo {author} {\bibfnamefont {W.}~\bibnamefont
  {Nazarewicz}},\ }\href {https://doi.org/10.1103/PhysRevC.51.R2880} {\bibfield
   {journal} {\bibinfo  {journal} {Phys. Rev. C}\ }\textbf {\bibinfo {volume}
  {51}},\ \bibinfo {pages} {R2880} (\bibinfo {year} {1995})}\BibitemShut
  {NoStop}%
\bibitem [{\citenamefont {Ryssens}\ \emph {et~al.}(2023)\citenamefont
  {Ryssens}, \citenamefont {Giacalone}, \citenamefont {Schenke},\ and\
  \citenamefont {Shen}}]{ryssens2023}%
  \BibitemOpen
  \bibfield  {author} {\bibinfo {author} {\bibfnamefont {W.}~\bibnamefont
  {Ryssens}}, \bibinfo {author} {\bibfnamefont {G.}~\bibnamefont {Giacalone}},
  \bibinfo {author} {\bibfnamefont {B.}~\bibnamefont {Schenke}},\ and\ \bibinfo
  {author} {\bibfnamefont {C.}~\bibnamefont {Shen}},\ }\href
  {https://doi.org/10.1103/PhysRevLett.130.212302} {\bibfield  {journal}
  {\bibinfo  {journal} {Phys. Rev. Lett.}\ }\textbf {\bibinfo {volume} {130}},\
  \bibinfo {pages} {212302} (\bibinfo {year} {2023})}\BibitemShut {NoStop}%
\bibitem [{\citenamefont {Meyer}\ \emph {et~al.}(2005)\citenamefont {Meyer},
  \citenamefont {Graw}, \citenamefont {Hertenberger}, \citenamefont {Wirth},
  \citenamefont {Casten}, \citenamefont {von Brentano}, \citenamefont
  {Bucurescu}, \citenamefont {Heinze}, \citenamefont {Jerke}, \citenamefont
  {Jolie}, \citenamefont {Krücken}, \citenamefont {Mahgoub}, \citenamefont
  {Pejovic}, \citenamefont {Möller}, \citenamefont {Mücher},\ and\
  \citenamefont {Scholl}}]{Meyer2005}%
  \BibitemOpen
  \bibfield  {author} {\bibinfo {author} {\bibfnamefont {D.~A.}\ \bibnamefont
  {Meyer}}, \bibinfo {author} {\bibfnamefont {G.}~\bibnamefont {Graw}},
  \bibinfo {author} {\bibfnamefont {R.}~\bibnamefont {Hertenberger}}, \bibinfo
  {author} {\bibfnamefont {H.-F.}\ \bibnamefont {Wirth}}, \bibinfo {author}
  {\bibfnamefont {R.~F.}\ \bibnamefont {Casten}}, \bibinfo {author}
  {\bibfnamefont {P.}~\bibnamefont {von Brentano}}, \bibinfo {author}
  {\bibfnamefont {D.}~\bibnamefont {Bucurescu}}, \bibinfo {author}
  {\bibfnamefont {S.}~\bibnamefont {Heinze}}, \bibinfo {author} {\bibfnamefont
  {J.~L.}\ \bibnamefont {Jerke}}, \bibinfo {author} {\bibfnamefont
  {J.}~\bibnamefont {Jolie}}, \bibinfo {author} {\bibfnamefont
  {R.}~\bibnamefont {Krücken}}, \bibinfo {author} {\bibfnamefont
  {M.}~\bibnamefont {Mahgoub}}, \bibinfo {author} {\bibfnamefont
  {P.}~\bibnamefont {Pejovic}}, \bibinfo {author} {\bibfnamefont
  {O.}~\bibnamefont {Möller}}, \bibinfo {author} {\bibfnamefont
  {D.}~\bibnamefont {Mücher}},\ and\ \bibinfo {author} {\bibfnamefont
  {C.}~\bibnamefont {Scholl}},\ }\href
  {https://doi.org/10.1088/0954-3899/31/10/003} {\bibfield  {journal} {\bibinfo
   {journal} {Journal of Physics G: Nuclear and Particle Physics}\ }\textbf
  {\bibinfo {volume} {31}},\ \bibinfo {pages} {S1399} (\bibinfo {year}
  {2005})}\BibitemShut {NoStop}%
\bibitem [{\citenamefont {Engel}\ and\ \citenamefont
  {Menéndez}(2017)}]{Engel2017}%
  \BibitemOpen
  \bibfield  {author} {\bibinfo {author} {\bibfnamefont {J.}~\bibnamefont
  {Engel}}\ and\ \bibinfo {author} {\bibfnamefont {J.}~\bibnamefont
  {Menéndez}},\ }\href {https://doi.org/10.1088/1361-6633/aa5bc5} {\bibfield
  {journal} {\bibinfo  {journal} {Reports on Progress in Physics}\ }\textbf
  {\bibinfo {volume} {80}},\ \bibinfo {pages} {046301} (\bibinfo {year}
  {2017})}\BibitemShut {NoStop}%
\bibitem [{\citenamefont {Egido}\ and\ \citenamefont
  {Robledo}(1992)}]{Egido1992}%
  \BibitemOpen
  \bibfield  {author} {\bibinfo {author} {\bibfnamefont {J.~L.}\ \bibnamefont
  {Egido}}\ and\ \bibinfo {author} {\bibfnamefont {L.~M.}\ \bibnamefont
  {Robledo}},\ }\href {https://doi.org/10.1016/0375-9474(92)90294-T} {\bibfield
   {journal} {\bibinfo  {journal} {Nuclear Physics A}\ }\textbf {\bibinfo
  {volume} {545}},\ \bibinfo {pages} {589} (\bibinfo {year}
  {1992})}\BibitemShut {NoStop}%
\bibitem [{\citenamefont {Robledo}(1994)}]{Rob94}%
  \BibitemOpen
  \bibfield  {author} {\bibinfo {author} {\bibfnamefont {L.~M.}\ \bibnamefont
  {Robledo}},\ }\href {https://doi.org/10.1103/PhysRevC.50.2874} {\bibfield
  {journal} {\bibinfo  {journal} {Physical Review C}\ }\textbf {\bibinfo
  {volume} {50}},\ \bibinfo {pages} {2874} (\bibinfo {year}
  {1994})}\BibitemShut {NoStop}%
\bibitem [{\citenamefont {Robledo}(2022{\natexlab{a}})}]{Robledo2022}%
  \BibitemOpen
  \bibfield  {author} {\bibinfo {author} {\bibfnamefont {L.~M.}\ \bibnamefont
  {Robledo}},\ }\href {https://doi.org/10.1103/PhysRevC.105.L021307} {\bibfield
   {journal} {\bibinfo  {journal} {Phys. Rev. C}\ }\textbf {\bibinfo {volume}
  {105}},\ \bibinfo {pages} {L021307} (\bibinfo {year}
  {2022}{\natexlab{a}})}\BibitemShut {NoStop}%
\bibitem [{\citenamefont {Robledo}(2022{\natexlab{b}})}]{Robledo2022a}%
  \BibitemOpen
  \bibfield  {author} {\bibinfo {author} {\bibfnamefont {L.~M.}\ \bibnamefont
  {Robledo}},\ }\href {https://doi.org/10.1103/PhysRevC.105.044317} {\bibfield
  {journal} {\bibinfo  {journal} {Phys. Rev. C}\ }\textbf {\bibinfo {volume}
  {105}},\ \bibinfo {pages} {044317} (\bibinfo {year}
  {2022}{\natexlab{b}})}\BibitemShut {NoStop}%
\bibitem [{\citenamefont {Robledo}(2010)}]{Robledo2010}%
  \BibitemOpen
  \bibfield  {author} {\bibinfo {author} {\bibfnamefont {L.~M.}\ \bibnamefont
  {Robledo}},\ }\href {https://doi.org/10.1088/0954-3899/37/6/064020}
  {\bibfield  {journal} {\bibinfo  {journal} {Journal of Physics G-nuclear and
  Particle Physics}\ }\textbf {\bibinfo {volume} {37}},\ \bibinfo {pages}
  {064020} (\bibinfo {year} {2010})}\BibitemShut {NoStop}%
\bibitem [{\citenamefont {Sheikh}\ \emph {et~al.}(2021)\citenamefont {Sheikh},
  \citenamefont {Dobaczewski}, \citenamefont {Ring}, \citenamefont {Robledo},\
  and\ \citenamefont {Yannouleas}}]{Sheikh2021}%
  \BibitemOpen
  \bibfield  {author} {\bibinfo {author} {\bibfnamefont {J.~A.}\ \bibnamefont
  {Sheikh}}, \bibinfo {author} {\bibfnamefont {J.}~\bibnamefont {Dobaczewski}},
  \bibinfo {author} {\bibfnamefont {P.}~\bibnamefont {Ring}}, \bibinfo {author}
  {\bibfnamefont {L.~M.}\ \bibnamefont {Robledo}},\ and\ \bibinfo {author}
  {\bibfnamefont {C.}~\bibnamefont {Yannouleas}},\ }\href
  {https://doi.org/10.1088/1361-6471/ac288a} {\bibfield  {journal} {\bibinfo
  {journal} {Journal of Physics G: Nuclear and Particle Physics}\ }\textbf
  {\bibinfo {volume} {48}},\ \bibinfo {pages} {123001} (\bibinfo {year}
  {2021})}\BibitemShut {NoStop}%
\bibitem [{\citenamefont {Rodriguez-Guzman}\ \emph {et~al.}(2012)\citenamefont
  {Rodriguez-Guzman}, \citenamefont {Robledo},\ and\ \citenamefont
  {Sarriguren}}]{RodriguezGuzman2012}%
  \BibitemOpen
  \bibfield  {author} {\bibinfo {author} {\bibfnamefont {R.}~\bibnamefont
  {Rodriguez-Guzman}}, \bibinfo {author} {\bibfnamefont {L.~M.}\ \bibnamefont
  {Robledo}},\ and\ \bibinfo {author} {\bibfnamefont {P.}~\bibnamefont
  {Sarriguren}},\ }\href {https://doi.org/10.1103/PhysRevC.86.034336}
  {\bibfield  {journal} {\bibinfo  {journal} {Physical Review C}\ }\textbf
  {\bibinfo {volume} {86}},\ \bibinfo {pages} {034336} (\bibinfo {year}
  {2012})}\BibitemShut {NoStop}%
\bibitem [{\citenamefont {Robledo}\ \emph {et~al.}(2019)\citenamefont
  {Robledo}, \citenamefont {Rodríguez},\ and\ \citenamefont
  {Rodríguez-Guzmán}}]{Robledo2019}%
  \BibitemOpen
  \bibfield  {author} {\bibinfo {author} {\bibfnamefont {L.~M.}\ \bibnamefont
  {Robledo}}, \bibinfo {author} {\bibfnamefont {T.~R.}\ \bibnamefont
  {Rodríguez}},\ and\ \bibinfo {author} {\bibfnamefont {R.~R.}\ \bibnamefont
  {Rodríguez-Guzmán}},\ }\href
  {http://stacks.iop.org/0954-3899/46/i=1/a=013001} {\bibfield  {journal}
  {\bibinfo  {journal} {Journal of Physics G: Nuclear and Particle Physics}\
  }\textbf {\bibinfo {volume} {46}},\ \bibinfo {pages} {013001} (\bibinfo
  {year} {2019})}\BibitemShut {NoStop}%
\bibitem [{\citenamefont {Berger}\ \emph {et~al.}(1984)\citenamefont {Berger},
  \citenamefont {Girod},\ and\ \citenamefont {Gogny}}]{berger1984}%
  \BibitemOpen
  \bibfield  {author} {\bibinfo {author} {\bibfnamefont {J.~F.}\ \bibnamefont
  {Berger}}, \bibinfo {author} {\bibfnamefont {M.}~\bibnamefont {Girod}},\ and\
  \bibinfo {author} {\bibfnamefont {D.}~\bibnamefont {Gogny}},\ }\href@noop {}
  {\bibfield  {journal} {\bibinfo  {journal} {Nucl. Phys. A}\ }\textbf
  {\bibinfo {volume} {428}},\ \bibinfo {pages} {23} (\bibinfo {year}
  {1984})}\BibitemShut {NoStop}%
\bibitem [{\citenamefont {Gonzalez-Boquera}\ \emph {et~al.}(2018)\citenamefont
  {Gonzalez-Boquera}, \citenamefont {Centelles}, \citenamefont {Viñas},\ and\
  \citenamefont {Robledo}}]{GonzalezBoquera2018}%
  \BibitemOpen
  \bibfield  {author} {\bibinfo {author} {\bibfnamefont {C.}~\bibnamefont
  {Gonzalez-Boquera}}, \bibinfo {author} {\bibfnamefont {M.}~\bibnamefont
  {Centelles}}, \bibinfo {author} {\bibfnamefont {X.}~\bibnamefont {Viñas}},\
  and\ \bibinfo {author} {\bibfnamefont {L.}~\bibnamefont {Robledo}},\ }\href
  {https://doi.org/https://doi.org/10.1016/j.physletb.2018.02.005} {\bibfield
  {journal} {\bibinfo  {journal} {Physics Letters B}\ }\textbf {\bibinfo
  {volume} {779}},\ \bibinfo {pages} {195 } (\bibinfo {year}
  {2018})}\BibitemShut {NoStop}%
\bibitem [{\citenamefont {Goriely}\ \emph {et~al.}(2009)\citenamefont
  {Goriely}, \citenamefont {Hilaire}, \citenamefont {Girod},\ and\
  \citenamefont {P\'{e}ru}}]{Goriely2009}%
  \BibitemOpen
  \bibfield  {author} {\bibinfo {author} {\bibfnamefont {S.}~\bibnamefont
  {Goriely}}, \bibinfo {author} {\bibfnamefont {S.}~\bibnamefont {Hilaire}},
  \bibinfo {author} {\bibfnamefont {M.}~\bibnamefont {Girod}},\ and\ \bibinfo
  {author} {\bibfnamefont {S.}~\bibnamefont {P\'{e}ru}},\ }\href@noop {}
  {\bibfield  {journal} {\bibinfo  {journal} {Phys. Rev. Lett.}\ }\textbf
  {\bibinfo {volume} {102}},\ \bibinfo {pages} {242501} (\bibinfo {year}
  {2009})}\BibitemShut {NoStop}%
\bibitem [{\citenamefont {Vinas}\ \emph {et~al.}(2019)\citenamefont {Vinas},
  \citenamefont {Gonzalez-Boquera}, \citenamefont {Centelles}, \citenamefont
  {Mondal},\ and\ \citenamefont {Robledo}}]{Vinas2019}%
  \BibitemOpen
  \bibfield  {author} {\bibinfo {author} {\bibfnamefont {X.}~\bibnamefont
  {Vinas}}, \bibinfo {author} {\bibfnamefont {C.}~\bibnamefont
  {Gonzalez-Boquera}}, \bibinfo {author} {\bibfnamefont {M.}~\bibnamefont
  {Centelles}}, \bibinfo {author} {\bibfnamefont {C.}~\bibnamefont {Mondal}},\
  and\ \bibinfo {author} {\bibfnamefont {L.}~\bibnamefont {Robledo}},\ }\href
  {https://www.actaphys.uj.edu.pl/S/12/3/705/pdf} {\bibfield  {journal}
  {\bibinfo  {journal} {Acta Physica Polonica B}\ }\textbf {\bibinfo {volume}
  {12}},\ \bibinfo {pages} {705} (\bibinfo {year} {2019})}\BibitemShut
  {NoStop}%
\bibitem [{\citenamefont {Bertsch}(1968)}]{Bertsch1968}%
  \BibitemOpen
  \bibfield  {author} {\bibinfo {author} {\bibfnamefont {G.}~\bibnamefont
  {Bertsch}},\ }\href
  {https://doi.org/https://doi.org/10.1016/0370-2693(68)90503-0} {\bibfield
  {journal} {\bibinfo  {journal} {Physics Letters B}\ }\textbf {\bibinfo
  {volume} {26}},\ \bibinfo {pages} {130} (\bibinfo {year} {1968})}\BibitemShut
  {NoStop}%
\bibitem [{\citenamefont {Hendrie}\ \emph {et~al.}(1968)\citenamefont
  {Hendrie}, \citenamefont {Glendenning}, \citenamefont {Harvey}, \citenamefont
  {Jarvis}, \citenamefont {Duhm}, \citenamefont {Saudinos},\ and\ \citenamefont
  {Mahoney}}]{Hendrie1968}%
  \BibitemOpen
  \bibfield  {author} {\bibinfo {author} {\bibfnamefont {D.}~\bibnamefont
  {Hendrie}}, \bibinfo {author} {\bibfnamefont {N.}~\bibnamefont
  {Glendenning}}, \bibinfo {author} {\bibfnamefont {B.}~\bibnamefont {Harvey}},
  \bibinfo {author} {\bibfnamefont {O.}~\bibnamefont {Jarvis}}, \bibinfo
  {author} {\bibfnamefont {H.}~\bibnamefont {Duhm}}, \bibinfo {author}
  {\bibfnamefont {J.}~\bibnamefont {Saudinos}},\ and\ \bibinfo {author}
  {\bibfnamefont {J.}~\bibnamefont {Mahoney}},\ }\href
  {https://doi.org/https://doi.org/10.1016/0370-2693(68)90502-9} {\bibfield
  {journal} {\bibinfo  {journal} {Physics Letters B}\ }\textbf {\bibinfo
  {volume} {26}},\ \bibinfo {pages} {127} (\bibinfo {year} {1968})}\BibitemShut
  {NoStop}%
\bibitem [{\citenamefont {Spieker}\ \emph {et~al.}(2023)\citenamefont
  {Spieker}, \citenamefont {Agbemava}, \citenamefont {Bazin}, \citenamefont
  {Biswas}, \citenamefont {Cottle}, \citenamefont {Farris}, \citenamefont
  {Gade}, \citenamefont {Ginter}, \citenamefont {Giraud}, \citenamefont
  {Kemper}, \citenamefont {Li}, \citenamefont {Nazarewicz}, \citenamefont
  {Noji}, \citenamefont {Pereira}, \citenamefont {Riley}, \citenamefont
  {Smith}, \citenamefont {Weisshaar},\ and\ \citenamefont
  {Zegers}}]{Spieker2023}%
  \BibitemOpen
  \bibfield  {author} {\bibinfo {author} {\bibfnamefont {M.}~\bibnamefont
  {Spieker}}, \bibinfo {author} {\bibfnamefont {S.}~\bibnamefont {Agbemava}},
  \bibinfo {author} {\bibfnamefont {D.}~\bibnamefont {Bazin}}, \bibinfo
  {author} {\bibfnamefont {S.}~\bibnamefont {Biswas}}, \bibinfo {author}
  {\bibfnamefont {P.}~\bibnamefont {Cottle}}, \bibinfo {author} {\bibfnamefont
  {P.}~\bibnamefont {Farris}}, \bibinfo {author} {\bibfnamefont
  {A.}~\bibnamefont {Gade}}, \bibinfo {author} {\bibfnamefont {T.}~\bibnamefont
  {Ginter}}, \bibinfo {author} {\bibfnamefont {S.}~\bibnamefont {Giraud}},
  \bibinfo {author} {\bibfnamefont {K.}~\bibnamefont {Kemper}}, \bibinfo
  {author} {\bibfnamefont {J.}~\bibnamefont {Li}}, \bibinfo {author}
  {\bibfnamefont {W.}~\bibnamefont {Nazarewicz}}, \bibinfo {author}
  {\bibfnamefont {S.}~\bibnamefont {Noji}}, \bibinfo {author} {\bibfnamefont
  {J.}~\bibnamefont {Pereira}}, \bibinfo {author} {\bibfnamefont
  {L.}~\bibnamefont {Riley}}, \bibinfo {author} {\bibfnamefont
  {M.}~\bibnamefont {Smith}}, \bibinfo {author} {\bibfnamefont
  {D.}~\bibnamefont {Weisshaar}},\ and\ \bibinfo {author} {\bibfnamefont
  {R.}~\bibnamefont {Zegers}},\ }\href
  {https://doi.org/https://doi.org/10.1016/j.physletb.2023.137932} {\bibfield
  {journal} {\bibinfo  {journal} {Physics Letters B}\ }\textbf {\bibinfo
  {volume} {841}},\ \bibinfo {pages} {137932} (\bibinfo {year}
  {2023})}\BibitemShut {NoStop}%
\bibitem [{Note1()}]{Note1}%
  \BibitemOpen
  \bibinfo {note} {By using the Gaussian Overlap approximation \cite
  {Ring1980}, the one dimensional GCM method can be approximated by a
  collective Schrodinger equation with a collective inertia given in terms of
  second derivatives of the Hamiltonian overlaps, see \cite {Ring1980} for
  details}\BibitemShut {NoStop}%
\bibitem [{\citenamefont {Ring}\ and\ \citenamefont {Schuck}(1980)}]{Ring1980}%
  \BibitemOpen
  \bibfield  {author} {\bibinfo {author} {\bibfnamefont {P.}~\bibnamefont
  {Ring}}\ and\ \bibinfo {author} {\bibfnamefont {P.}~\bibnamefont {Schuck}},\
  }\href@noop {} {\emph {\bibinfo {title} {The nuclear many body problem}}}\
  (\bibinfo  {publisher} {Springer-Verlag},\ \bibinfo {year}
  {1980})\BibitemShut {NoStop}%
\bibitem [{\citenamefont {Ryssens}\ \emph {et~al.}(2022)\citenamefont
  {Ryssens}, \citenamefont {Scamps}, \citenamefont {Goriely},\ and\
  \citenamefont {Bender}}]{Ryssens2022}%
  \BibitemOpen
  \bibfield  {author} {\bibinfo {author} {\bibfnamefont {W.}~\bibnamefont
  {Ryssens}}, \bibinfo {author} {\bibfnamefont {G.}~\bibnamefont {Scamps}},
  \bibinfo {author} {\bibfnamefont {S.}~\bibnamefont {Goriely}},\ and\ \bibinfo
  {author} {\bibfnamefont {M.}~\bibnamefont {Bender}},\ }\href
  {https://doi.org/10.1140/epja/s10050-022-00894-5} {\bibfield  {journal}
  {\bibinfo  {journal} {The European Physical Journal A}\ }\textbf {\bibinfo
  {volume} {58}},\ \bibinfo {pages} {246} (\bibinfo {year} {2022})}\BibitemShut
  {NoStop}%
\bibitem [{\citenamefont {Robledo}(2015)}]{Robledo2015}%
  \BibitemOpen
  \bibfield  {author} {\bibinfo {author} {\bibfnamefont {L.~M.}\ \bibnamefont
  {Robledo}},\ }\href {https://doi.org/10.1088/0954-3899/42/5/055109}
  {\bibfield  {journal} {\bibinfo  {journal} {Journal of Physics G: Nuclear and
  Particle Physics}\ }\textbf {\bibinfo {volume} {42}},\ \bibinfo {pages}
  {055109} (\bibinfo {year} {2015})}\BibitemShut {NoStop}%
\bibitem [{\citenamefont {{Brookhaven National Nuclear Data Center}}()}]{bnl}%
  \BibitemOpen
  \bibfield  {author} {\bibinfo {author} {\bibnamefont {{Brookhaven National
  Nuclear Data Center}}},\ }\href@noop {} {\bibinfo {title} {{ENSDF Data
  Base}}},\ \bibinfo {howpublished} {{http://www.nndc.bnl.gov}}\BibitemShut
  {NoStop}%
\bibitem [{\citenamefont {Garrett}(2001)}]{Garrett2001}%
  \BibitemOpen
  \bibfield  {author} {\bibinfo {author} {\bibfnamefont {P.~E.}\ \bibnamefont
  {Garrett}},\ }\href {https://doi.org/10.1088/0954-3899/27/1/201} {\bibfield
  {journal} {\bibinfo  {journal} {Journal of Physics G: Nuclear and Particle
  Physics}\ }\textbf {\bibinfo {volume} {27}},\ \bibinfo {pages} {R1} (\bibinfo
  {year} {2001})}\BibitemShut {NoStop}%
\bibitem [{\citenamefont {Garrett}\ \emph {et~al.}(2018)\citenamefont
  {Garrett}, \citenamefont {Wood},\ and\ \citenamefont {Yates}}]{Garrett2018}%
  \BibitemOpen
  \bibfield  {author} {\bibinfo {author} {\bibfnamefont {P.~E.}\ \bibnamefont
  {Garrett}}, \bibinfo {author} {\bibfnamefont {J.~L.}\ \bibnamefont {Wood}},\
  and\ \bibinfo {author} {\bibfnamefont {S.~W.}\ \bibnamefont {Yates}},\ }\href
  {https://doi.org/10.1088/1402-4896/aaba1c} {\bibfield  {journal} {\bibinfo
  {journal} {Physica Scripta}\ }\textbf {\bibinfo {volume} {93}},\ \bibinfo
  {pages} {063001} (\bibinfo {year} {2018})}\BibitemShut {NoStop}%
\bibitem [{\citenamefont {Giuliani}\ and\ \citenamefont
  {Robledo}(2018)}]{Giuliani2018}%
  \BibitemOpen
  \bibfield  {author} {\bibinfo {author} {\bibfnamefont {S.~A.}\ \bibnamefont
  {Giuliani}}\ and\ \bibinfo {author} {\bibfnamefont {L.~M.}\ \bibnamefont
  {Robledo}},\ }\href
  {https://doi.org/https://doi.org/10.1016/j.physletb.2018.10.045} {\bibfield
  {journal} {\bibinfo  {journal} {Physics Letters B}\ }\textbf {\bibinfo
  {volume} {787}},\ \bibinfo {pages} {134 } (\bibinfo {year}
  {2018})}\BibitemShut {NoStop}%
\bibitem [{\citenamefont {Lechaftois}\ \emph {et~al.}(2015)\citenamefont
  {Lechaftois}, \citenamefont {Deloncle},\ and\ \citenamefont
  {P\'eru}}]{Lechaftois2015}%
  \BibitemOpen
  \bibfield  {author} {\bibinfo {author} {\bibfnamefont {F.}~\bibnamefont
  {Lechaftois}}, \bibinfo {author} {\bibfnamefont {I.}~\bibnamefont
  {Deloncle}},\ and\ \bibinfo {author} {\bibfnamefont {S.}~\bibnamefont
  {P\'eru}},\ }\href {https://doi.org/10.1103/PhysRevC.92.034315} {\bibfield
  {journal} {\bibinfo  {journal} {Phys. Rev. C}\ }\textbf {\bibinfo {volume}
  {92}},\ \bibinfo {pages} {034315} (\bibinfo {year} {2015})}\BibitemShut
  {NoStop}%
\bibitem [{\citenamefont {Götz}\ \emph {et~al.}(1972)\citenamefont {Götz},
  \citenamefont {Pauli}, \citenamefont {Alder},\ and\ \citenamefont
  {Junker}}]{GOTZ19721}%
  \BibitemOpen
  \bibfield  {author} {\bibinfo {author} {\bibfnamefont {U.}~\bibnamefont
  {Götz}}, \bibinfo {author} {\bibfnamefont {H.}~\bibnamefont {Pauli}},
  \bibinfo {author} {\bibfnamefont {K.}~\bibnamefont {Alder}},\ and\ \bibinfo
  {author} {\bibfnamefont {K.}~\bibnamefont {Junker}},\ }\href
  {https://doi.org/https://doi.org/10.1016/0375-9474(72)90002-4} {\bibfield
  {journal} {\bibinfo  {journal} {Nuclear Physics A}\ }\textbf {\bibinfo
  {volume} {192}},\ \bibinfo {pages} {1} (\bibinfo {year} {1972})}\BibitemShut
  {NoStop}%
\bibitem [{\citenamefont {Erb}\ \emph {et~al.}(1972)\citenamefont {Erb},
  \citenamefont {Holden}, \citenamefont {Lee}, \citenamefont {Saladin},\ and\
  \citenamefont {Saylor}}]{Erb1972}%
  \BibitemOpen
  \bibfield  {author} {\bibinfo {author} {\bibfnamefont {K.~A.}\ \bibnamefont
  {Erb}}, \bibinfo {author} {\bibfnamefont {J.~E.}\ \bibnamefont {Holden}},
  \bibinfo {author} {\bibfnamefont {I.~Y.}\ \bibnamefont {Lee}}, \bibinfo
  {author} {\bibfnamefont {J.~X.}\ \bibnamefont {Saladin}},\ and\ \bibinfo
  {author} {\bibfnamefont {T.~K.}\ \bibnamefont {Saylor}},\ }\href
  {https://doi.org/10.1103/PhysRevLett.29.1010} {\bibfield  {journal} {\bibinfo
   {journal} {Phys. Rev. Lett.}\ }\textbf {\bibinfo {volume} {29}},\ \bibinfo
  {pages} {1010} (\bibinfo {year} {1972})}\BibitemShut {NoStop}%
\bibitem [{\citenamefont {Ronningen}\ \emph {et~al.}(1977)\citenamefont
  {Ronningen}, \citenamefont {Hamilton}, \citenamefont {Varnell}, \citenamefont
  {Lange}, \citenamefont {Ramayya}, \citenamefont {Garcia-Bermudez},
  \citenamefont {Lourens}, \citenamefont {Riedinger}, \citenamefont {McGowan},
  \citenamefont {Stelson}, \citenamefont {Robinson},\ and\ \citenamefont
  {Ford}}]{Ronningen1977}%
  \BibitemOpen
  \bibfield  {author} {\bibinfo {author} {\bibfnamefont {R.~M.}\ \bibnamefont
  {Ronningen}}, \bibinfo {author} {\bibfnamefont {J.~H.}\ \bibnamefont
  {Hamilton}}, \bibinfo {author} {\bibfnamefont {L.}~\bibnamefont {Varnell}},
  \bibinfo {author} {\bibfnamefont {J.}~\bibnamefont {Lange}}, \bibinfo
  {author} {\bibfnamefont {A.~V.}\ \bibnamefont {Ramayya}}, \bibinfo {author}
  {\bibfnamefont {G.}~\bibnamefont {Garcia-Bermudez}}, \bibinfo {author}
  {\bibfnamefont {W.}~\bibnamefont {Lourens}}, \bibinfo {author} {\bibfnamefont
  {L.~L.}\ \bibnamefont {Riedinger}}, \bibinfo {author} {\bibfnamefont {F.~K.}\
  \bibnamefont {McGowan}}, \bibinfo {author} {\bibfnamefont {P.~H.}\
  \bibnamefont {Stelson}}, \bibinfo {author} {\bibfnamefont {R.~L.}\
  \bibnamefont {Robinson}},\ and\ \bibinfo {author} {\bibfnamefont {J.~L.~C.}\
  \bibnamefont {Ford}},\ }\href {https://doi.org/10.1103/PhysRevC.16.2208}
  {\bibfield  {journal} {\bibinfo  {journal} {Phys. Rev. C}\ }\textbf {\bibinfo
  {volume} {16}},\ \bibinfo {pages} {2208} (\bibinfo {year}
  {1977})}\BibitemShut {NoStop}%
\end{thebibliography}
%

\end{document}